\documentclass[twocolumn,aps,pra,longbibliography]{revtex4-1}
\pdfoutput=1

\usepackage{amsmath,amssymb,bm,color,graphicx}

\newcommand{\real}{\mathrm{Re}}
\newcommand{\imag}{\mathrm{Im}}
\newcommand{\uone}{\bm{I}}
\newcommand{\tr}{\mathrm{tr}}
\newcommand{\invar}{\bm{\mathcal{J}}}
\newcommand{\axial}{\bm{\mathrm{axial}}}

\begin{document}

\title{Rotating Gaussian wave packets in weak external potentials}

\author{Arseni Goussev}

\affiliation{Department of Mathematics, Physics and Electrical Engineering, Northumbria University, Newcastle Upon Tyne NE1 8ST, United Kingdom}

\date{\today}

\begin{abstract}
	We address the time evolution of two- and three-dimensional nonrelativistic Gaussian wave packets in the presence of a weak external potential of arbitrary functional form. The focus of our study is the phenomenon of rotation of a Gaussian wave packet around its center of mass, as quantified by mean angular momentum computed relative to the wave-packet center. Using a semiclassical approximation of the eikonal type, we derive an explicit formula for a time-dependent change of mean angular momentum of a wave packet induced by its interaction with a weak external potential. As an example, we apply our analytical approach to the scenario of a two-dimensional quantum particle crossing a tilted ridge potential barrier. In particular, we demonstrate that the initial orientation of the particle wave packet determines the sense of its rotation, and report a good agreement between analytical and numerical results.
\end{abstract}


\maketitle

\section{Introduction}

Among many motivations to study the time evolution of quantum matter-wave packets two are particularly noteworthy.
First, localized wave packets provide the most natural tool for investigating the correspondence between quantum and classical motion. Indeed, while the center of a propagating wave packet traces a trajectory, a concept essential in classical mechanics, its finite spatial extent makes quantum interference effects possible.
Second, any initial state of a quantum system can be represented as a superposition of a number, finite or infinite, of localized wave packets. This fact, along with the linearity of quantum evolution, ensures that one's ability to predict the motion of each individual wave packet offers a way to quantitatively describe the time-evolution of an arbitrary, often complex, initial state.
Despite a large body of literature on quantum wave packet dynamics, much of it reviewed in Refs.~\cite{Lit86semiclassical, Hel91Wavepacket, YU00Physics, Rob04Quantum, Hel06Guided, Tan07Introduction}, the subject is by no means exhausted; many stimulating studies have appeared in recent years \cite{MYS10Correlations, VGH+10Optimally, GS11Wave, SVT12How, HF12Nonentangling, SA13Superluminal, DD14Transmission, Dod15Rotating, Sok15Interference, BBBB15Time, Dod16Rotating, PSC+16Poissons, PVT16Generalized, KK16Semiclassical, Str17Recovery}.

Of particular interest to the present work is a recent paper by Dodonov \cite{Dod15Rotating}, in which the author addresses the time evolution of nonrelativistic two-dimensional Gaussian wave packets possessing a finite value of mean angular momentum (MAM) \footnote{The tomographic probability description of rotating Gaussian wave packets in two dimensions has been recently reported in Ref.~\cite{Zhe16Optical}.} \footnote{Rotating Gaussian wave packets are different from electron vortices, which typically have a ring-shaped form with zero probability density at the center (see Refs.~\cite{BBSN07Semiclassical, BDN11Relativistic, IS11Scattering, BSVN12Electron, GM12Propagation, Kar15Gaussian, LBG+16Nondestructive, BB17Relativistic, Bar17Relativistic}, and Ref.~\cite{BIG+Theory} for a recent review).}. The value is the sum of an ``external'' (classical) part, related to the motion of the center of mass of the wave packet, and an ``internal'' (quantum-mechanical) contribution, resulting from the rotation of the wave packet around its center of mass and being a signature of nonzero position-momentum correlation. Internal rotation of atomic clouds of Gaussian shape has been successfully realized in laboratory experiments \cite{CLT07Observation}. One of several interesting features of rotating Gaussians is the effect of initial shrinking of wave packets with large enough position-momentum correlation coefficients, a phenomenon that may potentially be used to improve precision of the electron microscopy \cite{Dod15Rotating}.

In this paper, we investigate the dynamics of rotating Gaussian wave packets in the presence of weak external potentials, i.e. potentials whose variations are small compared to the kinetic energy of the moving particle. Our focus is the dependence of the internal MAM on the propagation time. Using a semiclassical (short-wavelength) approximation to the full quantum-mechanical propagator, we obtain an explicit formula that gives the value of the internal MAM as a function of the propagation time, parameters of the initial wave packet and the external potential. We further demonstrate the effectiveness of our analytical approach by treating an example problem in which a two-dimensional quantum particle traverses a tilted ridge barrier, and show that the predictions given by our semiclassical formula agree with the full numerical solution.

The paper is organized as follows. In Sec.~\ref{sec:wp_motion}, we derive a semiclassical eikonal-type approximation for the time-dependent wave function of an $N$-dimensional nonrelativistic particle moving through a spatial region with a weak external potential of arbitrary functional form. Section~\ref{sec:mam} is devoted to a calculation of MAM. First, in Sec.~\ref{sec:mam-Gaussian}, we obtain a coordinate-independent expression for the internal MAM of a Gaussian wave packet. Then, in Sec.~\ref{sec:mam-eikonal}, we find an approximate expression for the internal MAM of a wave packet evolving in the presence of a weak potential. An example physical scenario in which a two-dimensional ($N=2$) particle traverses a tilted ridge potential barrier is considered in Sec.~\ref{sec:example}. We summarize our results and provide conclusions in Sec.~\ref{sec:summary} of the paper. All conceptually straightforward, but technically strenuous calculations are deferred to the Appendixes.

\section{Propagation of Gaussian wave packets in weak external potentials}
\label{sec:wp_motion}

We consider an $N$-dimensional spinless nonrelativistic quantum particle of mass $\mu$, whose wave function is given by
\begin{align}
	&\psi_{\bm{q}, \bm{v}, \bm{\Omega}}(\bm{x}) = \left( \frac{\mu}{\pi \hbar} \, \right)^{N/4} \left( \det \bm{\Omega}_{\imag} \right)^{1/4} \nonumber \\
	&\;\: \times \exp \left\{ i \frac{\mu}{\hbar} \left[ \frac{1}{2} (\bm{x} - \bm{q})^{\mathrm{T}} \bm{\Omega} (\bm{x} - \bm{q}) + \bm{v}^{\mathrm{T}} (\bm{x} - \bm{q}) \right] \right\} \,.
\label{eq:Gaussian}
\end{align}
Here, $\bm{q}$ and $\bm{v}$ are real $N$-dimensional column vectors representing the average position and velocity of the particle, respectively, and $\bm{\Omega} = \bm{\Omega}_{\real} + i \bm{\Omega}_{\imag}$ is a complex symmetric $N \times N$ matrix with a positive-definite imaginary part $\bm{\Omega}_{\imag}$. (Throughout the paper, the real and imaginary parts of any quantity $\mathcal{Z}$ are interchangeably denoted both by $\mathcal{Z}_{\real}$ and $\mathcal{Z}_{\imag}$ and by $\real(\mathcal{Z})$ and $\imag(\mathcal{Z})$, respectively.) We note that the positivity of all eigenvalues of $\bm{\Omega}_{\imag}$ guarantees that $\bm{\Omega}$ is an invertible matrix~\footnote{Complex matrices with positive-definite imaginary part, such as $\bm{\Omega}$, are called strictly dissipative. Their determinant is nonzero, and the matrices are nonsingular (invertable). See Ref.~\cite{Lon81note} and references within for further details.}.

\subsection{Free particle motion}

If the particle, initially described by Eq.~(\ref{eq:Gaussian}), evolved in free space, its wave function at time $t$ would be given by
\begin{equation}
	\Psi_0 (\bm{x},t) = \int_{\mathbb{R}^N} d^N \bm{y} \, K_0(\bm{x}-\bm{y}, t) \psi_{\bm{q}, \bm{v}, \bm{\Omega}}(\bm{y}) \,,
\label{eq:Psi0_1}
\end{equation}
where
\begin{equation}
	K_0(\bm{\xi}, \tau) = \left( \frac{\mu}{2 \pi i \hbar \tau} \right)^{N/2} \exp \left( i \frac{\mu |\bm{\xi}|^2}{2 \hbar \tau} \right)
\label{eq:K0}
\end{equation}
is the free-particle propagator in $N$ dimensions. A direct evaluation of the Gaussian integral in Eq.~(\ref{eq:Psi0_1}) yields (see, e.g., Ref.~\cite{Hel00Semiclassical})
\begin{equation}
	\Psi_0 (\bm{x},t) = e^{i \varphi'} \psi_{\bm{q}', \bm{v}, \bm{\Omega}'}(\bm{x}) \,,
\label{eq:Psi0_2}
\end{equation}
where the new position $\bm{q}'$ of the particle is determined by
\begin{equation}
	\bm{q}' = \bm{q} + \bm{v} t \,,
\label{eq:q-prime}
\end{equation}
the new complex matrix $\bm{\Omega}'$, quantifying the shape and position-momentum correlation of the wave packet, is given by
\begin{equation}
	(\bm{\Omega}')^{-1} = \bm{\Omega}^{-1} + \uone t \,,
\label{eq:Omega-prime}
\end{equation}
with $\uone$ denoting the identity matrix, and the time-dependent phase $\varphi'$ reads
\begin{equation}
	\varphi' = \frac{\mu |\bm{v}|^2}{2 \hbar} t + \frac{1}{2} \arg \big( \det (\uone - \bm{\Omega}' t) \big) \,.
\end{equation}
We note that the phase can equivalently be expressed as $\varphi' = \frac{\mu |\bm{v}|^2}{2 \hbar} t - \frac{1}{2} \arg \big( \det (\uone + \bm{\Omega} t) \big)$.

\subsection{Eikonal-type approximation}

In the case that the particle moves in the presence of a potential $V(\bm{x})$, the initial wave packet evolves into
\begin{equation}
	\Psi (\bm{x},t) = \int_{\mathbb{R}^N} d^N \bm{y} \, K(\bm{x}, \bm{y}, t) \psi_{\bm{q}, \bm{v}, \bm{\Omega}}(\bm{y}) \,,
\label{eq:Psi_1}
\end{equation}
where $K(\bm{x}, \bm{y}, t)$ is the quantum propagator corresponding to the Hamiltonian $\frac{\bm{p} \cdot \bm{p}}{2 \mu} + V(\bm{x})$, with $\bm{p} = -i \hbar \frac{\partial}{\partial \bm{x}}$ being the momentum operator. In what follows, we use the Van Vleck--Gutzwiller approximation to the true quantum propagator, given by \cite{Gut71Periodic, BM72Semiclassical, Gut90Chaos, Sto99Quantum}
\begin{align}
	K(\bm{x},\bm{y},t) \simeq
	&\left( \frac{1}{2 \pi i \hbar} \right)^{N/2} \sum_{\gamma} \left| \det \left( \frac{\partial^2 S_{\gamma}(\bm{x}, \bm{y}, t)}{\partial \bm{x} \partial \bm{y}} \right) \right|^{1/2} \nonumber \\
	&\times \exp \left( \frac{i}{\hbar} S_{\gamma}(\bm{x}, \bm{y}, t) - i \frac{\pi \nu_{\gamma}}{2} \right) \,.
\label{eq:Van_Vleck}
\end{align}
Here, the sum runs over all classical trajectories $\gamma$ leading from $\bm{y}$ to $\bm{x}$ in time $t$. More precisely, $\gamma$ labels a position-space path $\bm{r}(\tau)$ that satisfies Newton's equation $\mu \frac{d^2 \bm{r}}{d \tau^2} + \frac{\partial V(\bm{r})}{\partial \bm{r}} = 0$, along with the boundary conditions $\bm{r}(0) = \bm{y}$ and $\bm{r}(t) = \bm{x}$. The function $S_{\gamma}$ is the Hamilton's principle function along the trajectory $\gamma$, i.e.
\begin{equation}
	S_{\gamma}(\bm{x}, \bm{y}, t) = \int_0^t d\tau \left( \frac{\mu}{2} \left| \frac{d \bm{r}(\tau)}{d \tau} \right|^2 - V \big( \bm{r}(\tau) \big) \right) \,.
\end{equation}
Finally, $\nu_{\gamma}$ is the Maslov index that counts the number of conjugate points, including possible multiplicities, along trajectory $\gamma$.

Let us now consider a situation in which the magnitude of potential $V(\bm{x})$ is small compared to kinetic energy $E_0$ of the corresponding classical particle, i.e.
\begin{equation}
	|V(\bm{x})| \ll E_0 = \frac{\mu |\bm{v}|^2}{2}
\end{equation}
for all $\bm{x}$. In this case, we can assume that there is only one classical trajectory $\gamma$ connecting point $\bm{y}$ at time $0$ and point $\bm{x}$ at time $t$. Moreover, to the leading order in $|V|/E_0$, this trajectory can be approximated by a straight free-flight path $\bm{r}(\tau) = (\tau / t) \bm{x} + (1 - \tau / t) \bm{y}$ (see the appendix in Ref.~\cite{GR13Scattering} for a discussion of the basis for this approximation). Then, Hamilton's principle function, evaluated along the straight path, reads
\begin{equation}
	S_\gamma(\bm{x}, \bm{y}, t) = \frac{\mu |\bm{x} - \bm{y}|^2}{2 t} - U_{\bm{x}, \bm{y}} t \,,
\label{eq:S_free_path}
\end{equation}
where
\begin{equation}
	U_{\bm{x}, \bm{y}} = U_{\bm{y}, \bm{x}} = \int_0^1 d\alpha \, V\big( \alpha \bm{x} + (1-\alpha) \bm{y} \big) \,.
\label{eq:U_xy-def}
\end{equation}
Substituting Eq.~(\ref{eq:S_free_path}) into Eq.~(\ref{eq:Van_Vleck}), taking into account the fact that the sum involves a single trajectory and that the corresponding Maslov index equals zero, and keeping only the leading order contribution to the stability factor $\det \left( \partial^2 S_{\gamma} / \partial \bm{x} \partial \bm{y} \right)$, we obtain
\begin{equation}
	K(\bm{x}, \bm{y}, t) \simeq K_0(\bm{x}-\bm{y}, t) \exp \left( -\frac{i}{\hbar} U_{\bm{x}, \bm{y}} t \right) \,,
\label{eq:K_eikonal}
\end{equation}
where $K_0$ is given by Eq.~(\ref{eq:K0}).

The propagator given by Eq.~(\ref{eq:K_eikonal}) represents a time-dependent version of the eikonal approximation to high-energy scattering \cite{Sch56Approximation, Gla59High, Sak94Modern}. Here, the external potential $V$ is regarded as a weak perturbation that does not affect the underlying classical dynamics, and so does not ``bend'' the straight trajectory connecting points $\bm{y}$ and $\bm{x}$ in time $t$, but only adds an extra phase to the corresponding quantum probability amplitude. A similar approximation has been previously used in semiclassical studies of the Loschmidt echo in chaotic systems \cite{JP01Environment, CT02Sensitivity, CT03uniform, STB03Hypersensitivity, VH03Semiclassical, CPJ04Universality, Van06Dephasing}. A particular case of the eikonal propagator, Eq.~(\ref{eq:K_eikonal}), corresponding to the case of $V(\bm{x})$ being a weak Gaussian potential in two dimensions ($N=2$), was analyzed in Ref.~\cite{GR13Scattering}.

Substituting Eqs.~(\ref{eq:Gaussian}) and (\ref{eq:K_eikonal}) into Eq.~(\ref{eq:Psi_1}), we have
\begin{widetext}
\begin{equation}
	\Psi(\bm{x}, t) \simeq \left( \frac{\mu}{\pi \hbar} \, \right)^{N/4} \left( \det \bm{\Omega}_{\imag} \right)^{1/4} \left( \frac{\mu}{2 \pi i \hbar t} \right)^{N/2} \int_{\mathbb{R}^N} d^N \bm{\xi} \, \exp \left[ i \frac{\mu}{\hbar} \left( \frac{\bm{\xi}^{\mathrm{T}} \bm{\Omega} \bm{\xi}}{2} + \bm{v}^{\mathrm{T}} \bm{\xi} \right) + i \frac{\mu |\bm{x} - \bm{q} - \bm{\xi}|^2}{2 \hbar t} - \frac{i t}{\hbar} U_{\bm{x}, \bm{q} + \bm{\xi}} \right] \,.
\label{eq:Psi_2}
\end{equation}
We now assume that the position extent of the initial wave packet is small compared to the characteristic length scale of the potential. This allows us to approximate $U_{\bm{x}, \bm{q}+\bm{\xi}}$ by a second degree polynomial in $\bm{\xi}$, i.e.
\begin{equation}
	U_{\bm{x}, \bm{q}+\bm{\xi}} \simeq U_{\bm{x}, \bm{q}} + \left( \frac{\partial U_{\bm{x}, \bm{q}}}{\partial \bm{q}} \right)^{\mathrm{T}} \bm{\xi} + \frac{1}{2} \bm{\xi}^{\mathrm{T}} \frac{\partial^2 U_{\bm{x}, \bm{q}}}{\partial \bm{q}^2} \bm{\xi} \,.
\label{eq:U_expansion}
\end{equation}
Hereinafter, $\frac{\partial}{\partial \bm{q}}$ and $\frac{\partial}{\partial \bm{q}'}$ represent gradient column vectors, while successive application of two gradients produces a square matrix. For example, $\frac{\partial U_{\bm{x}, \bm{q}}}{\partial \bm{q}}$ is a column vector with $j^{\mathrm{th}}$ element given by $\left. \frac{\partial U_{\bm{x}, \bm{y}}}{\partial y_j} \right|_{\bm{y} = \bm{q}}$, and $\frac{\partial^2 U_{\bm{q}', \bm{q}}}{\partial \bm{q}' \partial \bm{q}}$ is a square matrix whose $jk^{\mathrm{th}}$ element equals $\left. \frac{\partial^2 U_{\bm{x}, \bm{y}}}{\partial x_k \partial y_j} \right|_{(\bm{x}, \bm{y}) = (\bm{q}', \bm{q})}$. Substituting Eq.~(\ref{eq:U_expansion}) into Eq.~(\ref{eq:Psi_2}), we get
\begin{align}
	\Psi(\bm{x}, t)
	&\simeq \left( \frac{\mu}{\pi \hbar} \, \right)^{N/4} \left( \det \bm{\Omega}_{\imag} \right)^{1/4} \left( \frac{\mu}{2 \pi i \hbar t} \right)^{N/2} \exp \left( i \frac{\mu |\bm{x} - \bm{q}|^2}{2 \hbar t} - i \frac{U_{\bm{x}, \bm{q}}}{\hbar} t \right) \nonumber \\
	& \phantom{=} \times \int_{\mathbb{R}^N} d^N \bm{\xi} \, \exp \left[ i \frac{\mu}{2 \hbar t} \bm{\xi}^{\mathrm{T}} \left( \uone + \bm{\Omega} t - \frac{1}{\mu} \frac{\partial^2 U_{\bm{x}, \bm{q}}}{\partial \bm{q}^2} t^2 \right) \bm{\xi} - i \frac{\mu}{\hbar t} \left( \bm{x} - \bm{q} - \bm{v} t + \frac{1}{\mu} \frac{\partial U_{\bm{x}, \bm{q}}}{\partial \bm{q}} t^2 \right)^{\mathrm{T}} \bm{\xi} \right] \,.
\label{eq:Psi_3}
\end{align}
\end{widetext}
Evaluating the $N$-dimensional Gaussian integral (see Appendix~\ref{app:derivation_of_Psi} for details) we obtain
\begin{equation}
	\Psi(\bm{x}, t) \simeq e^{i \hat{\varphi}} \psi_{\hat{\bm{q}}, \hat{\bm{v}}, \hat{\bm{\Omega}}}(\bm{x}) \,,
\label{eq:Psi_5}
\end{equation}
where $\psi$ is defined in Eq.~(\ref{eq:Gaussian}), and
\begin{equation}
	\hat{\bm{v}} = \bm{v} - \frac{1}{\mu} \frac{\partial U_{\bm{x}, \bm{q}}}{\partial \bm{q}} t \,,
\label{eq:v-hat}
\end{equation}
\begin{equation}
	\hat{\bm{q}} = \bm{q} + \hat{\bm{v}} t \,,
	\label{eq:q-hat}
\end{equation}
\begin{equation}
	\hat{\bm{\Omega}}^{-1} = \left( \bm{\Omega} - \frac{1}{\mu} \frac{\partial^2 U_{\bm{x}, \bm{q}}}{\partial \bm{q}^2} t \right)^{-1} + \uone t \,,
\label{eq:Omega-hat}
\end{equation}
and
\begin{equation}
	\hat{\varphi} = \frac{1}{\hbar} \left( \frac{\mu |\hat{\bm{v}}|^2}{2} - U_{\bm{x}, \bm{q}} \right) t + \frac{1}{2} \arg \big( \det (\uone - \hat{\bm{\Omega}} t ) \big) \,.
\label{eq:varphi-hat}
\end{equation}
As a consistency check, we note that in the limiting case of $V(\bm{x}) = 0$ wave function $\Psi(\bm{x}, t)$, predicted by Eq.~(\ref{eq:Psi_5}), coincides with free-particle wave packet $\Psi_0(\bm{x}, t)$, given by Eq.~(\ref{eq:Psi0_2}). We also point out that, in general, wave packet $\Psi(\bm{x}, t)$ does {\it not} have a Gaussian shape, the reason being that quantities $\hat{\bm{q}}$, $\hat{\bm{v}}$, $\hat{\bm{\Omega}}$, and $\hat{\varphi}$ may exhibit a complicated dependence on $\bm{x}$.

\section{Mean angular momentum}
\label{sec:mam}

We now address the time-dependence of mean angular momentum (MAM) of a moving wave packet. First, we derive a general coordinate-independent expression for MAM of two- or three-dimensional Gaussian wave packets, given by Eq.~(\ref{eq:Gaussian}), and then find an approximation for MAM of an eikonal wave packet, given by Eq.~(\ref{eq:Psi_5}).

\subsection{Gaussian wave packet}
\label{sec:mam-Gaussian}

MAM corresponding to wave function $\psi_{\bm{q}, \bm{v}, \bm{\Omega}}$, given by Eq.~(\ref{eq:Gaussian}) with $N = 2$ or $3$, is defined as
\begin{equation}
	\bm{\mathcal{L}}(\bm{q}, \bm{v}, \bm{\Omega}) = \int_{\mathbb{R}^N} d^N \bm{x} \, \psi_{\bm{q}, \bm{v}, \bm{\Omega}}^*(\bm{x}) (\bm{x} \times \bm{p}) \psi_{\bm{q}, \bm{v}, \bm{\Omega}}(\bm{x}) \,,
\end{equation}
with $\bm{p} = -i \hbar \frac{\partial}{\partial {\bm x}}$ being the momentum operator. A straightforward differentiation yields
\begin{equation}
	\bm{p} \psi_{\bm{q}, \bm{v}, \bm{\Omega}}(\bm{x}) = \mu \big[ \bm{v} + \bm{\Omega} (\bm{x} - \bm{q}) \big] \psi_{\bm{q}, \bm{v}, \bm{\Omega}}(\bm{x}) \,.
\end{equation}
Adopting the notation
\begin{equation}
	\langle \cdot \rangle = \int_{\mathbb{R}^N} d^N \bm{x} \, (\cdot) \, |\psi_{\bm{q}, \bm{v}, \bm{\Omega}} (\bm{x})|^2 \,,
\end{equation}
we write
\begin{align}
	\bm{\mathcal{L}}(\bm{q}, \bm{v}, \bm{\Omega})
	&= \mu \langle \bm{x} \times [ \bm{v} + \bm{\Omega} (\bm{x} - \bm{q}) ] \rangle \nonumber \\[0.2cm]
	&= \bm{\mathcal{L}}_{\mathrm{e}}(\bm{q}, \bm{v}) + \bm{\mathcal{L}}_{\mathrm{i}}(\bm{\Omega}) \,,
\label{eq:L_total}
\end{align}
where
\begin{equation}
	\bm{\mathcal{L}}_{\mathrm{e}}(\bm{q}, \bm{v}) = \mu \bm{q} \times \bm{v}
\label{eq:L_CM}
\end{equation}
is an ``external'' part of MAM of the wave packet related to the motion of its center of mass, and
\begin{align}
	\bm{\mathcal{L}}_{\mathrm{i}}(\bm{\Omega})
	&= \mu \langle (\bm{x} - \bm{q}) \times \bm{\Omega} (\bm{x} - \bm{q}) \rangle \nonumber \\[0.2cm]
	&= \mu \left( \frac{\mu}{\pi \hbar} \right)^{N/2} \sqrt{\det \bm{\Omega}_{\imag}} \nonumber \\
	&\phantom{=} \times \int_{\mathbb{R}^N}  d^N \bm{\xi} \, ( \bm{\xi} \times \bm{\Omega} \bm{\xi} ) \exp \left( -\frac{\mu}{\hbar} \bm{\xi}^{\mathrm{T}} \bm{\Omega}_{\imag} \bm{\xi} \right)
\end{align}
is a contribution associated with the ``internal'' rotation of the wave packet around the center of mass. Evaluating the last integral (see Appendix~\ref{app:derivation_of_L_R} for details of the calculation), we find
\begin{equation}
	\bm{\mathcal{L}}_{\mathrm{i}}(\bm{\Omega}) = \frac{\hbar}{2} \axial \big[ \bm{\Omega}_{\real} , (\bm{\Omega}_{\imag})^{-1} \big] \,.
\label{eq:L_R}
\end{equation}
Here, $[\cdot, \cdot]$ denotes a commutator. (Note that since $\bm{\Omega}$ is a symmetric matrix, $\big[ \bm{\Omega}_{\real} , (\bm{\Omega}_{\imag})^{-1} \big]$ is an antisymmetric matrix.) The operator $\axial$ gives the axial vector of an antisymmetric matrix (see, e.g., Ref.~\cite{GFA13Mechanics}): given a real antisymmetric matrix $\bm{M}$, $\bm{m} = \axial \, \bm{M}$ is a unique vector such that $\bm{M} \bm{u} = \bm{m} \times \bm{u}$ for any vector $\bm{u}$. Using the Levi-Civita symbol, $\epsilon_{jkl}$, the axial vector can be written as
\begin{equation}
	[\axial \, \bm{M}]_j = -\frac{1}{2} \sum_{kl} \epsilon_{jkl} M_{kl} \,,
\end{equation}
or, more explicitly,
\begin{equation}
	\axial \, \bm{M} = \left\{
	\begin{array}{ll}
		(0, 0, M_{21})^{\mathrm{T}} & \mathrm{if} \quad N = 2 \\[0.2cm]
		(M_{32}, M_{13}, M_{21})^{\mathrm{T}} & \mathrm{if} \quad N = 3
	\end{array} \right. \,.
\end{equation}

We conclude this subsection with the following three remarks. First, in view of identity
\begin{equation}
	\big[ \bm{\Omega}_{\real} , (\bm{\Omega}_{\imag})^{-1} \big] = -(\bm{\Omega}_{\imag})^{-1} [\bm{\Omega}_{\real} , \bm{\Omega}_{\imag}] (\bm{\Omega}_{\imag})^{-1} \,,
\end{equation}
it is clear that the internal MAM, $\bm{\mathcal{L}}_{\mathrm{i}}$, is nonzero if and only if the real and imaginary parts of the matrix $\bm{\Omega}$ do not commute with each other. This implies, for instance, that $\bm{\mathcal{L}}_{\mathrm{i}} = \bm{0}$ for radially symmetric Gaussian wave packets; in other words, in order for a wave packet to possess a nonzero internal MAM, it must have a shape of an ellipse in two dimensions or an ellipsoid in three dimensions. Second, in the case of a two-dimensional ($N=2$) Gaussian wave packet, the expression for $\bm{\mathcal{L}}_{\mathrm{i}}$ as given by Eq.~(\ref{eq:L_R}) is in agreement with the one obtained in Ref.~\cite{Dod15Rotating}. Third, as expected, MAM of a free-particle wave packet is conserved. In particular, one has $\bm{\mathcal{L}}_{\mathrm{e}}(\bm{q}', \bm{v}) = \bm{\mathcal{L}}_{\mathrm{e}}(\bm{q}, \bm{v})$ and $\bm{\mathcal{L}}_{\mathrm{i}}(\bm{\Omega}') = \bm{\mathcal{L}}_{\mathrm{i}}(\bm{\Omega})$, where $\bm{q}'$ and $\bm{\Omega}'$ are given by Eqs.~(\ref{eq:q-prime}) and (\ref{eq:Omega-prime}), respectively. The conservation of the internal MAM follows from the fact that
\begin{align}
	\big[ \bm{\Omega}'_{\real} , (\bm{\Omega}'_{\imag})^{-1} \big]
	&= \big[ \real \{ (\bm{\Omega}')^{-1} \} , \big( \imag \{ (\bm{\Omega}')^{-1} \} \big)^{-1} \big] \nonumber \\
	&= \big[ \real \{ \bm{\Omega}^{-1} \} + \uone t , \big( \imag \{ \bm{\Omega}^{-1} \} \big)^{-1} \big] \nonumber \\
	&= \big[ \real \{ \bm{\Omega}^{-1} \} , \big( \imag \{ \bm{\Omega}^{-1} \} \big)^{-1} \big] \,,
\end{align}
which shows that the commutator is independent of time.

\subsection{Eikonal wave packet}
\label{sec:mam-eikonal}

Now, we derive an approximate expression for the internal MAM of an eikonal wave packet, given by Eq.~(\ref{eq:Psi_5}). It proves convenient to start from the following representation of the eikonal wave packet [cf. Eq.~(\ref{eq:app:Psi_1})]:
\begin{equation}
	\Psi(\bm{x}, t) = \left( \frac{\mu}{\pi \hbar} \, \right)^{N/4} \left( \det \bm{\Omega}_{\imag} \right)^{1/4} \exp \left( -\frac{\Phi}{2} \right) \,,
\end{equation}
where
\begin{align}
	\Phi
	&= -i \frac{\mu}{\hbar} \bm{\chi}^{\mathrm{T}} \hat{\bm{\Omega}} \bm{\chi} - i \frac{2 \mu}{\hbar} \hat{\bm{v}}^{\mathrm{T}} \bm{\chi} - i \frac{\mu}{\hbar} |\hat{\bm{v}}|^2 t + i \frac{2}{\hbar} U_{\bm{x}, \bm{q}} t \nonumber \\
	&\phantom{=} + \ln \left[ \det \left( \uone + \bm{\Omega} t - \frac{1}{\mu} \frac{\partial^2 U_{\bm{x}, \bm{q}}}{\partial \bm{q}^2} t^2 \right) \right] \,,
\label{eq:Phi_def}
\end{align}
with
\begin{equation}
	\bm{\chi} = \bm{x} - \hat{\bm{q}} \,.
	\label{eq:chi_def}
\end{equation}
The corresponding probability density reads
\begin{equation}
	|\Psi(\bm{x},t)|^2 = \left( \frac{\mu}{\pi \hbar} \, \right)^{N/2} \sqrt{\det \bm{\Omega}_{\imag}} \, \exp(-\Phi_{\real}) \,,
\end{equation}
with
\begin{align}
	\Phi_{\real}
	&= \frac{\mu}{\hbar} \bm{\chi}^{\mathrm{T}} \hat{\bm{\Omega}}_{\imag} \bm{\chi} \nonumber \\
	&+ \real \left\{ \ln \left[ \det \left( \uone + \bm{\Omega} t - \frac{1}{\mu} \frac{\partial^2 U_{\bm{x}, \bm{q}}}{\partial \bm{q}^2} t^2 \right) \right] \right\} \,.
\label{eq:Phi_real}
\end{align}

Assuming that the external potential is weak, we expect probability density $|\Psi(\bm{x},t)|^2$ to have a shape of a slightly distorted Gaussian centered at $\bm{x} = \bm{x}_{\max}$, with $\bm{x}_{\max}$ determined by the following system of $N$ equations:
\begin{equation}
	\frac{\partial \Phi_{\real}}{\partial \bm{x}} \bigg|_{\bm{x} = \bm{x}_{\max}} = \bm{0} \,.
\label{eq:eqs_for_r_def}
\end{equation}
These equations are equivalent to (see Appendix~\ref{app:derivatives_of_Phi} for details of the calculation)
\begin{align}
	\Bigg\{ &\frac{\mu}{t^2} \big[ \hat{\bm{\Omega}}_{\imag} \bm{\chi} \big]_j + \left( \frac{\partial}{\partial x_j} \frac{\partial U_{\bm{x}, \bm{q}}}{\partial \bm{q}} \right)^{\mathrm{T}} \hat{\bm{\Omega}}_{\imag} \bm{\chi} \nonumber \\
	&+ \bm{\chi}^{\mathrm{T}} \left( \uone - \hat{\bm{\Omega}}_{\real} t \right) \left( \frac{\partial}{\partial x_j} \frac{\partial^2 U_{\bm{x}, \bm{q}}}{\partial \bm{q}^2} \right) \hat{\bm{\Omega}}_{\imag} \bm{\chi} \nonumber \\
	&- \frac{\hbar}{2 \mu} \tr \left[ \left( \uone - \hat{\bm{\Omega}}_{\real} t \right) \frac{\partial}{\partial x_j} \frac{\partial^2 U_{\bm{x}, \bm{q}}}{\partial \bm{q}^2} \right] \Bigg\} \Bigg|_{\bm{x} = \bm{x}_{\max}} = 0 \,,
\label{eq:eqs_for_r}
\end{align}
with $1 \le j \le N$. Treating external potential $V$ as being $\epsilon$-small, we look for the solution to Eq.~(\ref{eq:eqs_for_r}) in the form $\bm{x}_{\max} = \bm{x}^{(0)}_{\max} + \mathcal{O}(\epsilon)$. In the leading order, Eq.~(\ref{eq:eqs_for_r}) reads $\bm{\Omega}'_{\imag} \left( \bm{x}^{(0)}_{\max} - \bm{q}' \right) = \bm{0}$, where $\bm{q}'$ and $\bm{\Omega}'$ are defined in Eqs.~(\ref{eq:q-prime}) and (\ref{eq:Omega-prime}), respectively. Since matrix $\bm{\Omega}'_{\imag}$ is positive definite, the only solution to this equation is $\bm{x}^{(0)}_{\max} = \bm{q}'$. Hence,
\begin{equation}
	\bm{x}_{\max} = \bm{q}' + \mathcal{O}(\epsilon) \,.
\end{equation}

Having determined the position of the probability density peak, $\bm{x}_{\max}$, we compute Hessian $\bm{\Omega}_{\mathrm{eff}}$ of $i \hbar \Phi / 2 \mu$ at $\bm{x} = \bm{x}_{\max}$, i.e.
\begin{align}
	\bm{\Omega}_{\mathrm{eff}}
	&= \frac{i \hbar}{2 \mu} \frac{\partial^2 \Phi}{\partial \bm{x}^2} \bigg|_{\bm{x} = \bm{x}_{\max}} \nonumber \\[0.2cm]
	&= \frac{\partial^2}{\partial \bm{x}^2} \Bigg\{ \frac{1}{2} \bm{\chi}^{\mathrm{T}} \hat{\bm{\Omega}} \bm{\chi} + \hat{\bm{v}}^{\mathrm{T}} \bm{\chi} + \frac{1}{2} |\hat{\bm{v}}|^2 t - \frac{1}{\mu} U_{\bm{x}, \bm{q}} t \nonumber \\
	&\phantom{=} + \frac{i \hbar}{2 \mu} \ln \left[ \det \left( \uone + \bm{\Omega} t - \frac{1}{\mu} \frac{\partial^2 U_{\bm{x}, \bm{q}}}{\partial \bm{q}^2} t^2 \right) \right] \Bigg\} \Bigg|_{\bm{x} = \bm{x}_{\max}} \,.
\label{eq:Omega_eff}
\end{align}
Matrix $\bm{\Omega}_{\mathrm{eff}}$ determines the shape and position-momentum correlation of the best Gaussian approximation to the eikonal wave packet, $\Psi(\bm{x}, t)$, around the maximum of its probability density. Evaluating the partial derivatives in the last expression (see Appendix~\ref{app:derivation_of_Hessian} for details), we find
\begin{equation}
	\bm{\Omega}_{\mathrm{eff}} = \bm{\Omega}' + \Delta \bm{\Omega} + \mathcal{O}(\epsilon^2) \,,
\label{eq:Hessian}
\end{equation}
where
\begin{align}
	\Delta \bm{\Omega} = -\frac{t}{\mu} \bigg[
	&\invar \frac{\partial^2 U_{\bm{q}', \bm{q}}}{\partial \bm{q}^2} \invar + \invar \frac{\partial^2 U_{\bm{q}', \bm{q}}}{\partial \bm{q}' \partial \bm{q}} + \frac{\partial^2 U_{\bm{q}', \bm{q}}}{\partial \bm{q} \partial \bm{q}'} \invar \nonumber \\
	&+ \frac{\partial^2 U_{\bm{q}', \bm{q}}}{\partial \bm{q}'^2} + i \frac{\hbar t}{2 \mu} \frac{\partial^2}{\partial \bm{q}'^2} \tr \left( \invar \frac{\partial^2 U_{\bm{q}', \bm{q}}}{\partial \bm{q}^2} \right) \bigg]
	\label{eq:Delta_Omega}
\end{align}
and
\begin{equation}
	\invar = (\uone + \bm{\Omega} t)^{-1} = \uone - \bm{\Omega}' t \,.
\label{eq:J-def}
\end{equation}
Note that, in general, $\frac{\partial^2 U_{\bm{q}', \bm{q}}}{\partial \bm{q}' \partial \bm{q}} \not= \frac{\partial^2 U_{\bm{q}', \bm{q}}}{\partial \bm{q} \partial \bm{q}'}$ (consider, e.g., $U_{\bm{x}, \bm{y}} = x_1 y_2$), but $\frac{\partial^2 U_{\bm{q}', \bm{q}}}{\partial \bm{q}' \partial \bm{q}} = \left( \frac{\partial^2 U_{\bm{q}', \bm{q}}}{\partial \bm{q} \partial \bm{q}'} \right)^{\mathrm{T}}$.

The internal MAM associated with wave function $\Psi(\bm{x}, t)$ can now be approximated by
\begin{align}
	&\bm{\mathcal{L}}_{\mathrm{i}}(\bm{\Omega}_{\mathrm{eff}}) \nonumber \\
	&\quad = \frac{\hbar}{2} \axial \big[ \bm{\Omega}'_{\real} + \Delta \bm{\Omega}_{\real} , ( \bm{\Omega}'_{\imag} + \Delta \bm{\Omega}_{\imag} )^{-1} \big] + \mathcal{O}(\epsilon^2) \nonumber \\
	&\quad = \bm{\mathcal{L}}_{\mathrm{i}}(\bm{\Omega}) + \Delta \bm{\mathcal{L}}_{\mathrm{i}} + \mathcal{O}(\epsilon^2) \,,
\end{align}
where function $\bm{\mathcal{L}}_{\mathrm{i}}$ is defined in Eq.~(\ref{eq:L_R}), and the leading order change in the internal MAM, resulting from the particle-potential interaction, is given by
\begin{align}
	\Delta \bm{\mathcal{L}}_{\mathrm{i}} = \frac{\hbar}{2}
	&\axial \Big\{ \big[ \Delta \bm{\Omega}_{\real} , (\bm{\Omega}'_{\imag})^{-1} \big] \nonumber \\
	&+ \big[ (\bm{\Omega}'_{\imag})^{-1} \Delta \bm{\Omega}_{\imag} (\bm{\Omega}'_{\imag})^{-1}, \bm{\Omega}'_{\real} \big] \Big\} \,.
\label{eq:Delta_L_R}
\end{align}
In deriving the last expression, we have used
\begin{align}
	&( \bm{\Omega}'_{\imag} + \Delta \bm{\Omega}_{\imag} )^{-1} \nonumber \\
	&\quad = (\bm{\Omega}'_{\imag})^{-1} - (\bm{\Omega}'_{\imag})^{-1} \Delta \bm{\Omega}_{\imag} (\bm{\Omega}'_{\imag})^{-1} + \mathcal{O}(\epsilon^2) \,,
\end{align}
along with the fact that the axial vector of a sum of two matrices equals the sum of their axial vectors. Together, Eqs.~(\ref{eq:Delta_Omega}) and (\ref{eq:Delta_L_R}) express the change of the internal MAM of a moving wave packet induced by its interaction with an external potential; these formulas constitute the central analytical result of the present paper.

\section{Example system: Tilted ridge barrier in two dimensions}
\label{sec:example}

Here, we apply the analytical method developed in the previous section to an example physical system, and compare analytical predictions against the full numerical solution. The system consists of a two-dimensional wave packet traversing a localized potential in the shape of a tilted ridge barrier; the precise definition of the system is presented in the following subsection.

\subsection{Analytics}

We consider a two-dimensional ($N=2$) quantum particle that is initially described by wave packet $\psi_{\bm{q}, \bm{v}, \bm{\Omega}}$, given by Eq.~(\ref{eq:Gaussian}) with
\begin{equation}
	\bm{q} = \left(
	\begin{array}{c}
		-q \\[0.2cm]
		0
	\end{array} \right) \,, \qquad q > 0 \,,
\end{equation}
\begin{equation}
	\bm{v} = \left(
	\begin{array}{c}
		v \\[0.2cm]
		0
	\end{array} \right) \,, \qquad v > 0 \,,
\end{equation}
and
\begin{equation}
	\bm{\Omega} = i \left(
	\begin{array}{cc}
		\omega_1 & 0 \\[0.2cm]
		0 & \omega_2
	\end{array} \right) \,, \qquad \omega_1 > 0 \,, \quad \omega_2 > 0 \,.
\end{equation}
The diagonal form of $\bm{\Omega}$ implies that $\big[ \bm{\Omega}_{\real} , (\bm{\Omega}_{\imag})^{-1} \big] = \bm{0}$ and, consequently, that the internal MAM associated with the initial wave packet equals zero. The length scales $\sqrt{\frac{\hbar}{\mu \omega_1}}$ and $\sqrt{\frac{\hbar}{\mu \omega_2}}$ characterize the spatial extent of the wave packet in $x_1$ and $x_2$ directions, respectively. Here we assume that the size of the wave packet is small compared to the distance separating the wave packet center and the origin of the coordinate frame, i.e
\begin{equation}
	q^2 \gg \frac{\hbar}{\mu \omega_-} \,,
\label{eq:cond-small_wp}
\end{equation}
where $\omega_- = \min \{ \omega_1 , \omega_2 \}$.

In order to simplify the following calculations, we fix the propagation time to be
\begin{equation}
	t = \frac{2 q}{v} \,,
\label{eq:t-symmetric}
\end{equation}
so that the center of the free-particle wave packet at time $t$ is [cf. Eq.~(\ref{eq:q-prime})]
\begin{equation}
	\bm{q}' = \left(
	\begin{array}{c}
		q \\[0.2cm]
		0
	\end{array} \right) \,.
\end{equation}
The shape and position-momentum correlation of a free-particle wave packet at time $t$ is governed by Eq.~(\ref{eq:Omega-prime}), which, in the present case, reads
\begin{equation}
	\bm{\Omega}' = \frac{2 q}{v} \left(
	\begin{array}{cc}
		\omega_1^2 / \Xi_1 & 0 \\[0.2cm]
		0 & \omega_2^2 / \Xi_2
	\end{array} \right) + i \left(
	\begin{array}{cc}
		\omega_1 / \Xi_1 & 0 \\[0.2cm]
		0 & \omega_2 / \Xi_2
	\end{array} \right) \,,
\label{eq:Omega_prime-example}
\end{equation}
where
\begin{equation}
	\Xi_1 = 1 + \left( \frac{2 q}{v} \omega_1 \right)^2 \qquad \mathrm{and} \qquad \Xi_2 = 1 + \left( \frac{2 q}{v} \omega_2 \right)^2 \,.
\end{equation}
We further assume that the average velocity $v$ is sufficiently large for the size of the wave packet not to change substantially during propagation time $t$. In particular, we want the size of the wave packet at time $t$ to be small compared to the distance $q$ between its center and the origin. To this end, we assume
\begin{equation}
	v > 2 \omega_+ q \,,
\label{eq:cond-fast_wp}
\end{equation}
where $\omega_+ = \max \{ \omega_1, \omega_2 \}$.

The external potential has the form of a tilted ridge barrier:
\begin{equation}
	V(\bm{\xi}) = \xi_2 f(\xi_1) \,,
\label{eq:V}
\end{equation}
where $f$ is an even real-valued function, $f(-\xi) = f(\xi)$, that is localized on an interval of width $2 \ell$ centered around the origin, i.e.
\begin{equation}
	f(\xi) \simeq 0 \quad \mathrm{if} \quad |\xi| > \ell \,.
\end{equation}
The width of the potential barrier is assumed to be small compared to the distance between the wave packet and the origin:
\begin{equation}
	\ell \ll q \,.
\label{eq:cond-localized_potential}
\end{equation}
The parametric regime defined by Eqs.~(\ref{eq:cond-small_wp}), (\ref{eq:cond-fast_wp}) and (\ref{eq:cond-localized_potential}) implies that both the initial and propagated wave packets have a negligible overlap with the potential barrier.

In accordance with Eq.~(\ref{eq:U_xy-def}), we have
\begin{equation}
	U_{\bm{x}, \bm{y}} = \int_0^1 d\alpha \, \big( y_2 + (x_2 - y_2) \alpha \big) f \big( y_1 + (x_1 - y_1) \alpha \big) \,.
\label{eq:U_example}
\end{equation}
Since we are only interested in $U_{\bm{x}, \bm{y}}$ for $\bm{x}$ and $\bm{y}$ lying within a small (compared to $q$) vicinity of the points $\bm{q}'$ and $\bm{q}$, respectively, the integration interval in Eq.~(\ref{eq:U_example}) can be extended to the entire real axis. This gives
\begin{align}
	U_{\bm{x}, \bm{y}}
	&\simeq \int_{-\infty}^{+\infty} d\alpha \, \big( y_2 + (x_2 - y_2) \alpha \big) f \big( y_1 + (x_1 - y_1) \alpha \big) \nonumber \\[0.2cm]
	&= \frac{1}{x_1 - y_1} \int_{-\infty}^{+\infty} d\xi \, \left( \frac{x_1 y_2 - y_1 x_2}{x_1 - y_1} + \frac{x_2 - y_2}{x_1 - y_1} \xi \right) f(\xi) \,.
\end{align}
Using the fact that $f(\xi)$ is an even function, we obtain
\begin{equation}
	U_{\bm{x}, \bm{y}} \simeq V_0 \frac{x_1 y_2 - y_1 x_2}{(x_1 - y_1)^2} \,,
	\label{eq:U_tilted_ridge}
\end{equation}
where
\begin{equation}
	V_0 = \int_{-\infty}^{+\infty} d\xi \, f(\xi) \,.
\label{eq:V0_def}
\end{equation}

From Eq.~(\ref{eq:U_tilted_ridge}), we find by direct differentiation
\begin{equation}
	\frac{\partial^2 U_{\bm{q}',\bm{q}}}{\partial \bm{q}^2} = \frac{\partial^2 U_{\bm{x},\bm{y}}}{\partial \bm{y}^2} \bigg|_{(\bm{x},\bm{y})=(\bm{q}',\bm{q})} = \frac{V_0}{4 q^2} \left(
	\begin{array}{cc}
		0 & 1 \\[0.2cm]
		1 & 0
	\end{array} \right) \,,
\label{eq:example-matrix1}
\end{equation}
\begin{equation}
	\frac{\partial^2 U_{\bm{q}',\bm{q}}}{\partial \bm{q}' \partial \bm{q}} = \frac{\partial}{\partial \bm{x}} \frac{\partial U_{\bm{x},\bm{y}}}{\partial \bm{y}} \bigg|_{(\bm{x},\bm{y})=(\bm{q}',\bm{q})} = \bm{0} \,,
\label{eq:example-matrix2}
\end{equation}
\begin{equation}
	\frac{\partial^2 U_{\bm{q}',\bm{q}}}{\partial \bm{q} \partial \bm{q}'} = \frac{\partial}{\partial \bm{y}} \frac{\partial U_{\bm{x},\bm{y}}}{\partial \bm{x}} \bigg|_{(\bm{x},\bm{y})=(\bm{q}',\bm{q})} = \bm{0} \,,
\label{eq:example-matrix3}
\end{equation}
\begin{equation}
	\frac{\partial^2 U_{\bm{q}',\bm{q}}}{\partial \bm{q}'^2} = \frac{\partial^2 U_{\bm{x},\bm{y}}}{\partial \bm{x}^2} \bigg|_{(\bm{x},\bm{y})=(\bm{q}',\bm{q})} = -\frac{V_0}{4 q^2} \left(
	\begin{array}{cc}
		0 & 1 \\[0.2cm]
		1 & 0
	\end{array} \right) \,.
\label{eq:example-matrix4}
\end{equation}
Also, using Eq.~(\ref{eq:J-def}), we find
\begin{equation}
	\invar = \left(
	\begin{array}{cc}
		\frac{1}{1 + i 2 q \omega_1 / v} & 0 \\[0.2cm]
		0 & 	\frac{1}{1 + i 2 q \omega_2 / v}
	\end{array} \right) \,,
\label{eq:example-J}
\end{equation}
and, consequently,
\begin{align}
	\frac{\partial^2}{\partial \bm{q}'^2} &\tr \left( \invar \frac{\partial^2 U_{\bm{q}', \bm{q}}}{\partial \bm{q}^2} \right) \nonumber \\
	&= \frac{\partial^2}{\partial \bm{q}'^2} \left( \frac{1}{1 + i \frac{2 q}{v} \omega_1} \frac{\partial^2 U_{\bm{q}', \bm{q}}}{\partial q_1^2} + \frac{1}{1 + i \frac{2 q}{v} \omega_2} \frac{\partial^2 U_{\bm{q}', \bm{q}}}{\partial q_2^2} \right) \nonumber \\
	&= \left( \frac{1}{1 + i \frac{2 q}{v} \omega_1} \frac{\partial^2}{\partial q_1^2} + \frac{1}{1 + i \frac{2 q}{v} \omega_2 } \frac{\partial^2}{\partial q_2^2} \right) \frac{\partial^2 U_{\bm{q}', \bm{q}}}{\partial \bm{q}'^2} \,.
\end{align}
But since
\begin{equation}
	\frac{\partial^2}{\partial q_1^2} \frac{\partial^2 U_{\bm{q}', \bm{q}}}{\partial \bm{q}'^2} = \frac{\partial^2}{\partial y_1^2} \frac{\partial^2 U_{\bm{x}, \bm{y}}}{\partial \bm{x}^2} \bigg|_{(\bm{x},\bm{y})=(\bm{q}',\bm{q})} = \bm{0}
\end{equation}
and
\begin{equation}
	\frac{\partial^2}{\partial q_2^2} \frac{\partial^2 U_{\bm{q}', \bm{q}}}{\partial \bm{q}'^2} = \frac{\partial^2}{\partial y_2^2} \frac{\partial^2 U_{\bm{x}, \bm{y}}}{\partial \bm{x}^2} \bigg|_{(\bm{x},\bm{y})=(\bm{q}',\bm{q})} = \bm{0} \,,
\end{equation}
we obtain
\begin{equation}
	\frac{\partial^2}{\partial \bm{q}'^2} \tr \left( \invar \frac{\partial^2 U_{\bm{q}', \bm{q}}}{\partial \bm{q}^2} \right) = \bm{0} \,.
\label{eq:example-matrix5}
\end{equation}
We now have all the terms needed to calculate $\Delta \bm{\Omega}$. Indeed, substituting Eqs.~(\ref{eq:example-matrix1}--\ref{eq:example-J}) and (\ref{eq:example-matrix5}) into Eq.~(\ref{eq:Delta_Omega}), we find
\begin{equation}
	\Delta \bm{\Omega} = -\frac{V_0}{\mu v^2} \frac{\frac{2 q}{v} \omega_1 \omega_2 - i (\omega_1 + \omega_2)}{\left( 1 + i \frac{2 q}{v} \omega_1 \right) \left( 1 + i \frac{2 q}{v} \omega_2 \right)} \left(
	\begin{array}{cc}
		0 & 1 \\[0.2cm]
		1 & 0
	\end{array} \right) \,,
\end{equation}
and then
\begin{equation}
	\Delta \bm{\Omega}_{\real} = \frac{2 q V_0}{\mu v^3} \frac{\omega_1^2 + \omega_1 \omega_2 + \omega_2^2 + \left( \frac{2 q}{v} \omega_1 \omega_2 \right)^2}{\Xi_1 \Xi_2} \left(
	\begin{array}{cc}
		0 & 1 \\[0.2cm]
		1 & 0
	\end{array} \right)
\label{eq:Delta_Omega_real-example}
\end{equation}
and
\begin{equation}
	\Delta \bm{\Omega}_{\imag} = \frac{V_0}{\mu v^2} \frac{\omega_1 + \omega_2}{\Xi_1 \Xi_2} \left(
	\begin{array}{cc}
		0 & 1 \\[0.2cm]
		1 & 0
	\end{array} \right) \,.
\label{eq:Delta_Omega_imag-example}
\end{equation}
Using Eqs.~(\ref{eq:Omega_prime-example}) and (\ref{eq:Delta_Omega_imag-example}), we obtain
\begin{equation}
	(\bm{\Omega}'_{\imag})^{-1} \Delta \bm{\Omega}_{\imag} (\bm{\Omega}'_{\imag})^{-1} = \frac{V_0}{\mu v^2} \frac{\omega_1 + \omega_2}{\omega_1 \omega_2} \left(
	\begin{array}{cc}
		0 & 1 \\[0.2cm]
		1 & 0
	\end{array} \right) \,.
\label{eq:Delta_Omega_imag_mix-example}
\end{equation}
Finally, substituting Eqs.~(\ref{eq:Omega_prime-example}), (\ref{eq:Delta_Omega_real-example}) and (\ref{eq:Delta_Omega_imag_mix-example}) into Eq.~(\ref{eq:Delta_L_R}) and performing straightforward algebraic transformations, we obtain the following simple expression for the only nonzero (out-of-plane) component of the internal MAM at time~$t$:
\begin{equation}
	\big[ \Delta \bm{\mathcal{L}}_{\mathrm{i}} \big]_3 = \hbar \frac{q V_0}{\mu v^3} (\omega_1 - \omega_2) \,.
\label{eq:Delta_L_R-example}
\end{equation}

Equation~(\ref{eq:Delta_L_R-example}) prompts a number of interesting observations. First, the magnitude of final MAM rapidly decreases with increasing velocity of the particle. Second, initial wave packets that are circularly symmetric ($\omega_1 = \omega_2$) do not rotate after having interacted with the barrier. Third, the sense of rotation of a wave packet transmitted over the barrier is determined by its initial orientation. More precisely, the wave packets initially elongated in $x_2$ direction ($\omega_1 > \omega_2$) rotate in the positive sense upon crossing the barrier [see Fig.~\ref{fig:snapshots}(a,b)], while the wave packets initially elongated in $x_1$ direction ($\omega_1 < \omega_2$) acquire rotation in the negative sense [see Fig.~\ref{fig:snapshots}(c,d)]. 

\subsection{Numerics}

In order to estimate the accuracy of Eq.~(\ref{eq:Delta_L_R-example}), we simulate the wave packet motion numerically, taking $f$ to be a Gaussian function:
\begin{equation}
	f(\xi_1) = \frac{V_0}{\sqrt{\pi} \ell} e^{-(\xi_1 / \ell)^2} \,.
\label{eq:f-example}
\end{equation}
Note that this definition is consistent with Eq.~(\ref{eq:V0_def}). The parameters of the system are chosen as follows. The moving particle is taken to be a $^7$Li atom of mass $\mu = 7.016003 \,$u. The initial distance from the barrier is $q = 0.3 \,$mm, and the mean velocity is $v = 4 \,$mm/s. The final time of wave packet propagation, as given by Eq.~(\ref{eq:t-symmetric}), equals $t = 150\,$ms. The width of the barrier is taken to be $\ell = 20 \, \mu$m. In what follows, we investigate two initial scenarios: the initial wave packet is characterized (i) by $\omega_1 = 10 \,$s$^{-1}$ and $\omega_2 = 5 \,$s$^{-1}$, which corresponds to wave packet elongation along $x_2$ direction, and (ii) by $\omega_1 = 5 \,$s$^{-1}$ and $\omega_2 = 10 \,$s$^{-1}$, corresponding to elongation along $x_1$ direction. Frequency values $\omega_j = 10 \,$s$^{-1}$ and $5 \,$s$^{-1}$ correspond to the spatial wave packet extent of $\sqrt{\hbar / \mu \omega_j} \simeq 30.1\,\mu$m and $42.5\,\mu$m, respectively. We note that all the parameter values chosen above are comparable to typical values in modern atom-optics experiments \cite{FCG+11Realization, JMR+12Coherent, CFV+13Matter}. The wave packet propagation, for various values of the potential strength $V_0$, is simulated by expanding the full quantum-mechanical propagator in a series of Chebyshev polynomials of the Hamiltonian. A comprehensive description of the numerical method, along with implementation details, are given in Refs.~\cite{TK84accurate, RKMF03Unified, DBS11Efficiency}.

\begin{figure*}[htb!]
	\centering
	\includegraphics[width=0.7\textwidth]{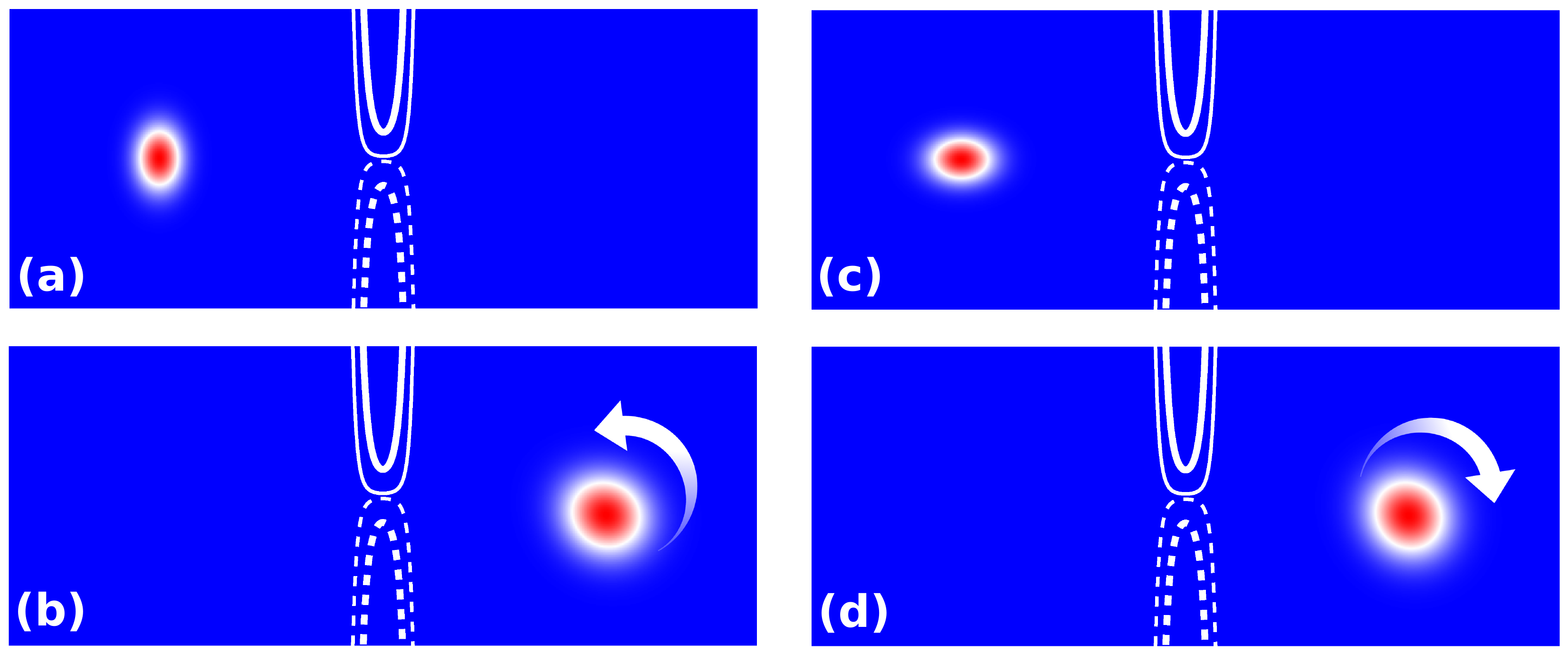}
	\caption{(Color online) Illustration of a Gaussian wave packet passing over the tilted ridge barrier, defined by Eqs.~(\ref{eq:V}) and (\ref{eq:f-example}). Four white curves show equipotential curves: the solid (dashed) curves correspond to positive (negative) values of the potential; the curve thickness increases as the absolute value of the potential becomes larger. The wave packet travels from left to right. Panels (a) and (b) show the initial and final probability densities, respectively, for the case of $\omega_1 > \omega_2$. Panels (c) and (d) are the corresponding pair of figures in the case of $\omega_1 < \omega_2$. The curly arrows in (b) and (d) show the sense of the internal rotation of the final wave packet.}
\label{fig:snapshots}
\end{figure*}

Figure~\ref{fig:snapshots} illustrates the qualitative behavior of the wave packet with $\omega_1 > \omega_2$, Fig.~\ref{fig:snapshots}(a,b), and with $\omega_1 < \omega_2$, Fig.~\ref{fig:snapshots}(c,d). In particular, the figure shows that, in agreement with Eq.~(\ref{eq:Delta_L_R-example}), the sense of final rotation of the wave packet depends on its initial orientation. The snapshots presented in the figure have been obtained from the numerical simulations with the parameter values given above and correspond to the potential strength of $V_0 = 10^{-13}\,$eV.

\begin{figure*}[htb!]
	\centering
	\includegraphics[width=0.7\textwidth]{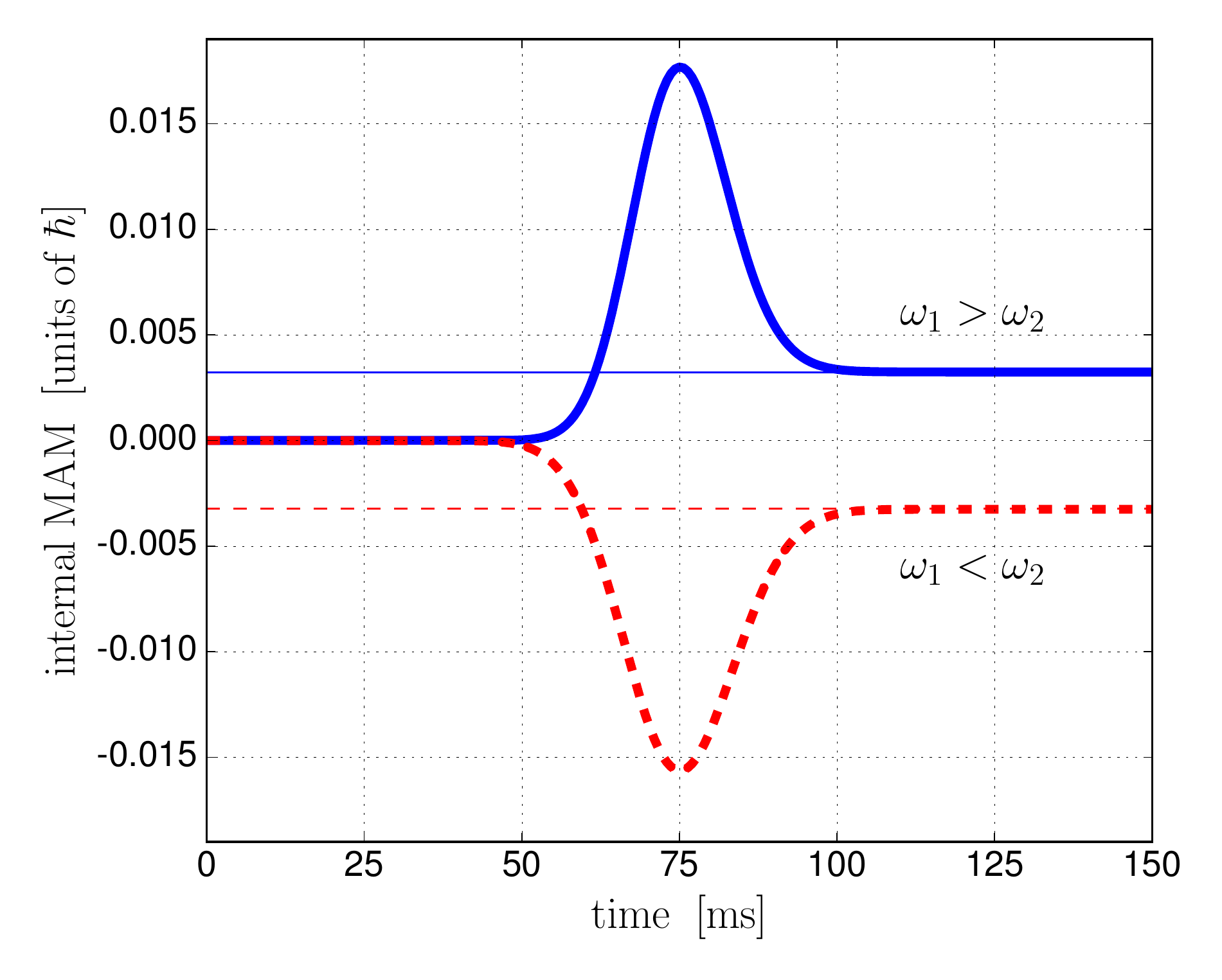}
	\caption{(Color online) Internal MAM (in units of $\hbar$) as a function of propagation time (in milliseconds) in a tilted ridge barrier system. The parameters of the system are as follows: $V_0 = 10^{-14} \,$eV, $\ell = 20 \, \mu$m, $\mu = 7.016003 \,$u ($^7$Li atom), $q = 0.3 \,$mm, and $v = 4 \,$mm/s. The (blue) thick solid curve represents the case of $\omega_1 = 10 \,$s$^{-1}$ and $\omega_2 = 5 \,$s$^{-1}$, and the (blue) thin solid line shows the corresponding value of $\big[ \Delta \bm{\mathcal{L}}_\mathrm{i} \big]_3$, as predicted by Eq.~(\ref{eq:Delta_L_R-example}). The (red) thick dashed curve represents the case of $\omega_1 = 5 \,$s$^{-1}$ and $\omega_2 = 10 \,$s$^{-1}$, and the (red) thin dashed line shows the corresponding value of $\big[ \Delta \bm{\mathcal{L}}_\mathrm{i} \big]_3$, as predicted by Eq.~(\ref{eq:Delta_L_R-example}).}
\label{fig:mam_vs_time}
\end{figure*}

Figure~\ref{fig:mam_vs_time} shows how MAM, computed with respect to the mean position of the moving wave packet, changes in time as the quantum particle crosses the tilted ridge barrier of strength $V_0 = 10^{-14}\,$eV. The (blue) thick solid curve represents the numerical results for the initial wave packet with $\omega_1 = 10 \,$s$^{-1}$ and $\omega_2 = 5 \,$s$^{-1}$, and the (red) thick dashed curve corresponds to the case of $\omega_1 = 5 \,$s$^{-1}$ and $\omega_2 = 10 \,$s$^{-1}$. Both curves exhibit well pronounced peaks of the absolute value of the internal MAM. It is interesting to observe that while the peaks are centered at the same instant of time, given by $q/v = 75\,$ms, their magnitudes are not the same. The thin solid (blue) and dashed (red) lines show the values of $\big[ \Delta \bm{\mathcal{L}}_\mathrm{i} \big]_3$, as predicted by Eq.~(\ref{eq:Delta_L_R-example}), for the cases of $\omega_1 > \omega_2$ and $\omega_1 < \omega_2$, respectively. It is clear that Eq.~(\ref{eq:Delta_L_R-example}) correctly approximates the value of internal MAM for the wave packet transmitted over the barrier.

\begin{figure*}[htb!]
	\centering
	\includegraphics[width=0.7\textwidth]{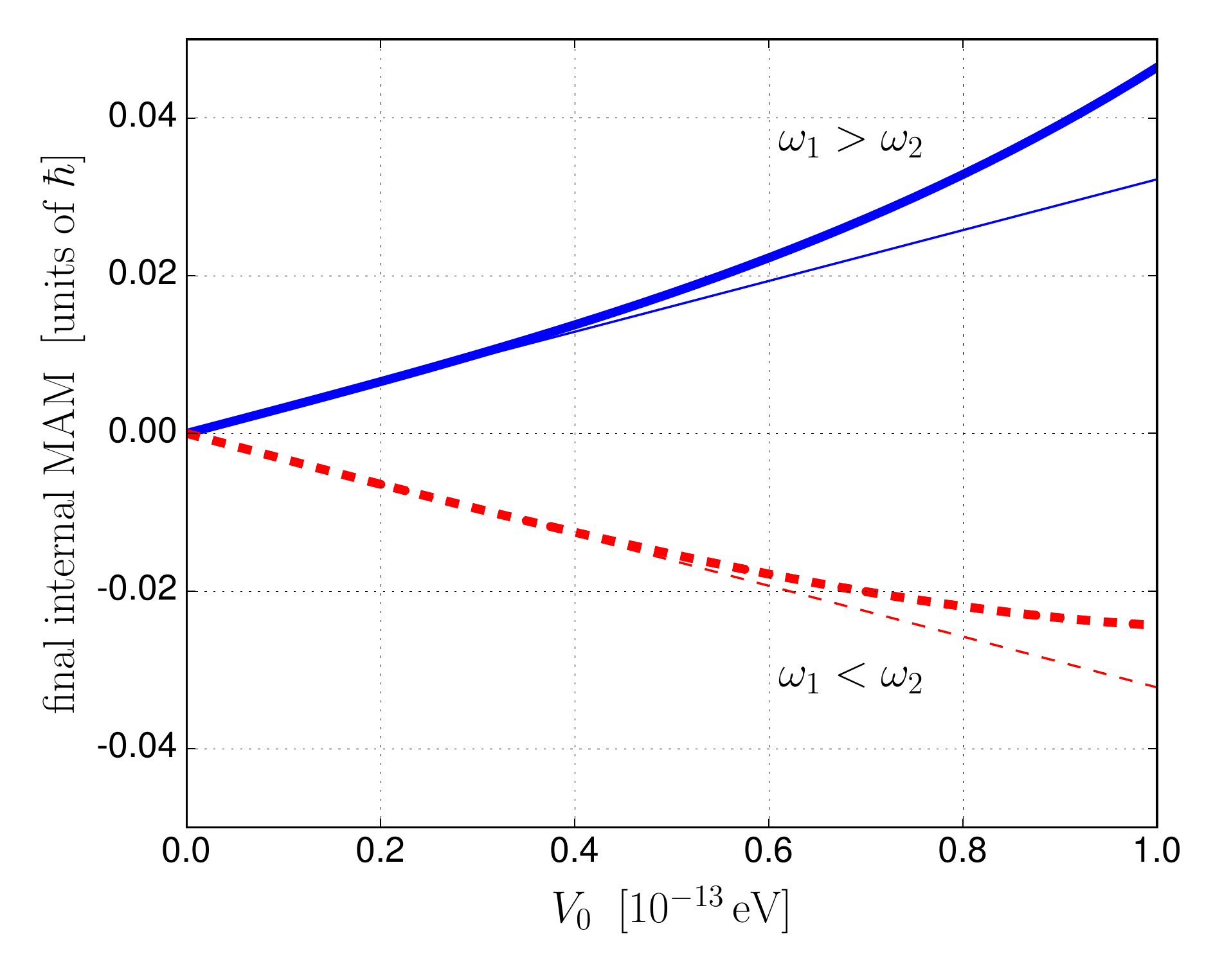}
	\caption{(Color online) Internal MAM (in units of $\hbar$) at time $t = 150\,$ms, at which the wave packet is located to the right of the barrier (cf.~Fig.~\ref{fig:mam_vs_time}), as a function of the potential strength (in units of $10^{-13}\,$eV). All parameter values are the same as in Fig.~\ref{fig:mam_vs_time}. The (blue) thick solid curve represents the case of $\omega_1 = 10 \,$s$^{-1}$ and $\omega_2 = 5 \,$s$^{-1}$, and the (blue) thin solid line shows the corresponding value of $\big[ \Delta \bm{\mathcal{L}}_\mathrm{i} \big]_3$, as predicted by Eq.~(\ref{eq:Delta_L_R-example}). The (red) thick dashed curve represents the case of $\omega_1 = 5 \,$s$^{-1}$ and $\omega_2 = 10 \,$s$^{-1}$, and the (red) thin dashed line shows the corresponding value of $\big[ \Delta \bm{\mathcal{L}}_\mathrm{i} \big]_3$, as predicted by Eq.~(\ref{eq:Delta_L_R-example}).}
\label{fig:mam_vs_V0}
\end{figure*}

Figure~\ref{fig:mam_vs_V0} provides a comparison between the final value of internal MAM obtained via numerical simulations and that predicted by Eq.~(\ref{eq:Delta_L_R-example}) for external potentials of various strengths. We find the analytical and numerical results to be in good agreement with each other for potentials of strength $|V_0| \lesssim 0.4 \times 10^{-13}\,$eV. This is consistent with the fact that our analytical calculations are only valid for sufficiently weak external potentials.

\section{Summary and conclusions}
\label{sec:summary}

In this paper we have analyzed the phenomenon of internal rotation of two- and three-dimensional quantum Gaussian wave packets in the presence of weak external potentials. The main outcome of our study is twofold. First, we have derived a simple coordinate-independent formula, Eq.~(\ref{eq:L_R}), that expresses the internal mean angular momentum of a Gaussian wave packet in terms of the commutator of the real and imaginary parts of the complex matrix determining the wave packet shape and position-momentum correlation. Second, using semiclassical analysis, we have obtained an explicit expression, given by Eqs.~(\ref{eq:Delta_Omega}) and (\ref{eq:Delta_L_R}), for the internal mean angular momentum of a Gaussian wave packet propagating through an arbitrary weak external potential. We have further tested our analytical method in the case of a two-dimensional wave packet crossing a tilted potential ridge barrier, finding analytical and numerical results to be in good agreement.

The analytical method presented in this paper has been obtained using a time-dependent semiclassical approximation of the eikonal type. The eikonal approximation is different from a more commonly used linear-dynamics approach, also known as the ``thawed'' Gaussian approximation (TGA) \cite{Hel75Time, Hel76Classical}, in which, at every time instant, the external potential is approximated by a second-degree polynomial around the center of the moving wave packet. There are two main differences between the eikonal and TGA methods. First, the TGA, unlike the eikonal approximation, assumes the spatial extent of the wave packet to be small compared to the characteristic length scale on which the potential varies \footnote{An extension of the TGA to the case of wide Gaussian wave packets has been proposed in Ref.~\cite{Gou11Nonmonotonic}.}. This imposes a natural limitation on the maximal propagation time, making the TGA a short-time asymptotic method. Second, the TGA requires for the center trajectory of the wave packet to be computed numerically. In contrast to the TGA and to the main advantage of the eikonal approximation, all calculations in the eikonal approximation can often be carried out analytically. However the downside of the eikonal approximation is that its applicability is limited to weak external potentials only, as evident from the example treated in Sec.~\ref{sec:example}. In the future, it would be interesting to make a systematic comparison between the effectiveness of the TGA and eikonal approximation in the problem of internal rotation of Gaussian wave packets.

The present study raises a number of interesting questions for future research. How does one extend the analytical results for the time-dependence of mean angular momentum presented in this paper to physical scenarios involving strong external potentials? For instance, how is the internal rotation of a Gaussian wave packet affected by strong scattering events, in which, e.g., the wave packet breaks into two or more spatially-separated parts? What is the upper bound on the internal angular momentum that a Gaussian wave packet may acquire in a scattering event? Is there an ``optimal'' external potential that would transfer the maximal amount of internal rotation to a Gaussian wave packet, while minimizing its shape distortions? These and other related questions address fundamental aspects of time-dependent quantum dynamics; we believe that any progress in this direction will improve our understanding of basic quantum physics and may subsequently lead to technological applications.

\begin{acknowledgments}
	The author thanks Ilya Arakelyan, Maximilien Barbier, Orestis Georgiou, and Roman Schubert for valuable comments and useful discussions, and acknowledges the financial support of EPSRC under Grant No.~EP/K024116/1.
\end{acknowledgments}

\appendix

\onecolumngrid

\section{Derivation of Eq.~(\ref{eq:Psi_5})}
\label{app:derivation_of_Psi}

For convenience, we introduce
\begin{equation}
	\bm{a}_{\bm{x}, \bm{q}} = \frac{\partial U_{\bm{x}, \bm{q}}}{\partial \bm{q}} \quad \mathrm{and} \quad \bm{B}_{\bm{x}, \bm{q}} = \frac{\partial^2 U_{\bm{x}, \bm{q}}}{\partial \bm{q}^2} \,.
	\label{eq:a_and_B}
\end{equation}
These definitions will also be used in Appendixes~\ref{app:derivatives_of_Phi} and \ref{app:derivation_of_Hessian}.

Using the integral identity (see, e.g., Ref.~\cite{Zin05Path})
\begin{equation}
	\int_{\mathbb{R}^N} d^N \bm{\xi} \, \exp \left(-\bm{\xi}^{\mathrm{T}} \bm{A} \bm{\xi} + \bm{b}^\mathrm{T} \bm{\xi} \right) = \sqrt{\frac{\pi^N}{\det \bm{A}}} \exp \left( \frac{1}{4} \bm{b}^\mathrm{T} \bm{A}^{-1} \bm{b} \right) \,,
	\label{eq:Gaussian_integral}
\end{equation}
with $\bm{A} = \frac{\mu}{2 i \hbar t} \left( \uone + \bm{\Omega} t - \frac{1}{\mu} \bm{B}_{\bm{x}, \bm{q}} t^2 \right)$ and $\bm{b} = \frac{\mu}{i \hbar t} \left( \bm{x} - \bm{q} - \bm{v} t + \frac{1}{\mu} \bm{a}_{\bm{x}, \bm{q}} t^2 \right)$, we rewrite Eq.~(\ref{eq:Psi_3}) as
\begin{align}
	\Psi(\bm{x}, t)
	&= \left( \frac{\mu}{\pi \hbar} \, \right)^{N/4} \frac{\left( \det \bm{\Omega}_{\imag} \right)^{1/4}}{\sqrt{\det \left( \uone + \bm{\Omega} t - \frac{1}{\mu} \bm{B}_{\bm{x}, \bm{q}} t^2 \right)}} \exp \Bigg\{ i \frac{\mu |\bm{x} - \bm{q}|^2}{2 \hbar t} - i \frac{U_{\bm{x}, \bm{q}}}{\hbar}  t \nonumber \\
	& \qquad - i \frac{\mu}{2 \hbar t} \left( \bm{x} - \bm{q} - \bm{v} t + \frac{1}{\mu} \bm{a}_{\bm{x}, \bm{q}} t^2 \right)^{\mathrm{T}} \left( \uone + \bm{\Omega} t - \frac{1}{\mu} \bm{B}_{\bm{x}, \bm{q}} t^2 \right)^{-1} \left( \bm{x} - \bm{q} - \bm{v} t + \frac{1}{\mu} \bm{a}_{\bm{x}, \bm{q}} t^2 \right) \Bigg\} \,.
	\label{eq:Psi_4}
\end{align}
Then, introducing
\begin{equation}
	\widetilde{\bm{\Omega}} = \bm{\Omega} - \frac{1}{\mu} \bm{B}_{\bm{x}, \bm{q}} t \,,
\label{eq:app:Omega_widetilde}
\end{equation}
and using the quantities $\hat{\bm{q}}$ and $\hat{\bm{v}}$, defined in Eqs.~(\ref{eq:q-hat}) and (\ref{eq:v-hat}), respectively, we obtain
\begin{align}
	\Psi(\bm{x}, t)
	&= \left( \frac{\mu}{\pi \hbar} \, \right)^{N/4} \frac{\left( \det \bm{\Omega}_{\imag} \right)^{1/4}}{\sqrt{\det ( \uone + \widetilde{\bm{\Omega}} t ) }} \exp \Bigg\{ i \frac{\mu |\bm{x} - \hat{\bm{q}} + \hat{\bm{v}} t|^2}{2 \hbar t} - i \frac{U_{\bm{x}, \bm{q}}}{\hbar} t - i \frac{\mu}{2 \hbar t} \left( \bm{x} - \hat{\bm{q}} \right)^{\mathrm{T}} \left( \uone + \widetilde{\bm{\Omega}} t \right)^{-1} \left( \bm{x} - \hat{\bm{q}} \right) \Bigg\} \nonumber \\[0.2cm]
	&= \left( \frac{\mu}{\pi \hbar} \, \right)^{N/4} \frac{\left( \det \bm{\Omega}_{\imag} \right)^{1/4}}{\sqrt{\det ( \uone + \widetilde{\bm{\Omega}} t ) }} \exp \Bigg\{ i \frac{\mu}{2 \hbar t} \left( \bm{x} - \hat{\bm{q}} \right)^{\mathrm{T}} \left[ \uone - \left( \uone + \widetilde{\bm{\Omega}} t \right)^{-1} \right] \left( \bm{x} - \hat{\bm{q}} \right) \nonumber \\
	& \hspace{5.3cm} + i \frac{\mu}{\hbar} \hat{\bm{v}}^{\mathrm{T}} \left( \bm{x} - \hat{\bm{q}} \right) + i \frac{\mu |\hat{\bm{v}}|^2}{2 \hbar} t - i \frac{U_{\bm{x}, \bm{q}}}{\hbar} t \Bigg\} \,.
\label{eq:app:Psi_1}
\end{align}
We then define
\begin{equation}
	\hat{\bm{\Omega}} = \frac{1}{t} \left[ \uone - \left( \uone + \widetilde{\bm{\Omega}} t \right)^{-1} \right] \,.
\label{eq:app:Omega-hat-def}
\end{equation}
This definition implies the equality
\begin{equation}
	\hat{\bm{\Omega}}^{-1} = \widetilde{\bm{\Omega}}^{-1} + \bm{I} t \,,
\label{eq:app:Omega-hat}
\end{equation}
which is equivalent to Eq.~(\ref{eq:Omega-hat}). This allows us to rewrite Eq.~(\ref{eq:app:Psi_1}) as
\begin{align}
	\Psi(\bm{x}, t)
	&= \left( \frac{\mu}{\pi \hbar} \, \right)^{N/4} \frac{\left( \det \bm{\Omega}_{\imag} \right)^{1/4}}{\sqrt{ \big| \det ( \uone + \widetilde{\bm{\Omega}} t ) \big| }} \exp \left[ i \frac{\mu}{2 \hbar} \left( \bm{x} - \hat{\bm{q}} \right)^{\mathrm{T}} \hat{\bm{\Omega}} \left( \bm{x} - \hat{\bm{q}} \right) + i \frac{\mu}{\hbar} \hat{\bm{v}}^{\mathrm{T}} \left( \bm{x} - \hat{\bm{q}} \right) + i \hat{\varphi} \right] \nonumber \\[0.2cm]
	&= \frac{\left( \det \bm{\Omega}_{\imag} \right)^{1/4}}{\sqrt{ \big| \det ( \uone + \widetilde{\bm{\Omega}} t ) \big| }} \, \frac{e^{i \hat{\varphi}} \psi_{\hat{\bm{q}}, \hat{\bm{v}}, \hat{\bm{\Omega}}}(\bm{x})}{\big( \det \hat{\bm{\Omega}}_{\imag} \big)^{1/4}} \,,
\label{eq:app:Psi_2}
\end{align}
where
\begin{equation}
	\hat{\varphi} = \frac{1}{\hbar} \left( \frac{\mu |\hat{\bm{v}}|^2}{2} - U_{\bm{x}, \bm{q}} \right) t - \frac{1}{2} \arg \big( \det (\uone + \widetilde{\bm{\Omega}} t ) \big) \,.
\label{eq:app:varphi}
\end{equation}
We note that Eq.~(\ref{eq:app:varphi}) is equivalent to Eq.~(\ref{eq:varphi-hat}). Indeed, it follows from Eq.~(\ref{eq:app:Omega-hat-def}) that
\begin{equation}
	\uone + \widetilde{\bm{\Omega}} t = \left( \uone - \hat{\bm{\Omega}} t \right)^{-1} \,,
\label{eq:app:tilde-hat-identity}
\end{equation}
which, in turn, implies that $\arg \big( \det (\uone + \widetilde{\bm{\Omega}} t ) \big) = \arg \big( 1 / \det (\uone - \hat{\bm{\Omega}} t ) \big) = -\arg \big( \det (\uone - \hat{\bm{\Omega}} t ) \big)$. Then, since the external potential is real, so is the matrix $\bm{B}_{\bm{x}, \bm{q}}$. This implies $\bm{\Omega}_{\imag} = \widetilde{\bm{\Omega}}_{\imag}$. Also, from Eq.~(\ref{eq:app:Omega-hat}) it follow that $\uone + \widetilde{\bm{\Omega}} t = \hat{\bm{\Omega}}^{-1} \widetilde{\bm{\Omega}}$. Taking the last two facts into account, we rewrite Eq.~(\ref{eq:app:Psi_2}) as
\begin{equation}
	\Psi(\bm{x}, t) = \left( \frac{\det \widetilde{\bm{\Omega}}_{\imag}}{\big| \det \widetilde{\bm{\Omega}} \big|^2} \, \frac{\big| \det \hat{\bm{\Omega}} \big|^2}{\det \hat{\bm{\Omega}}_{\imag}} \right)^{1/4} e^{i \hat{\varphi}} \psi_{\hat{\bm{q}}, \hat{\bm{v}}, \hat{\bm{\Omega}}}(\bm{x})
\end{equation}
It now remains to prove that
\begin{equation}
	\frac{\det \hat{\bm{\Omega}}_{\imag}}{\big| \det \hat{\bm{\Omega}} \big|^2} = \frac{\det \widetilde{\bm{\Omega}}_{\imag}}{\big| \det \widetilde{\bm{\Omega}} \big|^2} \,.
\label{eq:app:identity}
\end{equation}
To this end, we rewrite the left-hand side of the last equality as
\begin{align}
	\frac{\det \hat{\bm{\Omega}}_{\imag}}{\big| \det \hat{\bm{\Omega}} \big|^2}
	&= \frac{\det \left( \tfrac{1}{2 i} (\hat{\bm{\Omega}} - \hat{\bm{\Omega}}^*) \right)}{ ( \det  \hat{\bm{\Omega}} ) ( \det  \hat{\bm{\Omega}}^* ) } \nonumber \\[0.2cm]
	&= \det \left( \tfrac{1}{2 i} \hat{\bm{\Omega}}^{-1} (\hat{\bm{\Omega}} - \hat{\bm{\Omega}}^*) (\hat{\bm{\Omega}}^*)^{-1} \right) \nonumber \\[0.2cm]
	&= \det \left( -\tfrac{1}{2 i} \big[ \hat{\bm{\Omega}}^{-1} - (\hat{\bm{\Omega}}^*)^{-1} \big] \right) \nonumber \\[0.2cm]
	&= \det \left( -\imag ( \hat{\bm{\Omega}}^{-1} ) \right) \,.
\label{eq:app:LHS}
\end{align}
Similarly, the right-hand side of the equality reads
\begin{equation}
	\frac{\det \widetilde{\bm{\Omega}}_{\imag}}{\big| \det \widetilde{\bm{\Omega}} \big|^2} = \det \left( -\imag ( \widetilde{\bm{\Omega}}^{-1} ) \right) \,.
\label{eq:app:RHS}
\end{equation}
Finally, since $\imag ( \hat{\bm{\Omega}}^{-1} ) = \imag ( \widetilde{\bm{\Omega}}^{-1} + \uone t) = \imag ( \widetilde{\bm{\Omega}}^{-1} )$, the right-hand sides of Eqs.~(\ref{eq:app:LHS}) and (\ref{eq:app:RHS}) are equal to one another, which concludes the proof of Eq.~(\ref{eq:app:identity}).

\section{Derivation of Eq.~(\ref{eq:L_R})}
\label{app:derivation_of_L_R}

Here we show that
\begin{equation}
	\int_{\mathbb{R}^N} d^N \bm{x} \, ( \bm{x} \times \bm{R} \bm{x} ) \exp \left( -\bm{x}^{\mathrm{T}} \bm{S} \bm{x} \right) = \frac{1}{2} \sqrt{\frac{\pi^N}{\det \bm{S}}} \, \axial \big[ \bm{R}, \bm{S}^{-1} \big] \,,
\label{eq:app:axial_integral}
\end{equation}
where $\bm{S}$ is a real symmetric positive-definite matrix, and $\bm{R}$ is a real symmetric matrix. Once this identity is proven, Eq.~(\ref{eq:L_R}) follows immediately.

We start by diagonalizing $\bm{S}$:
\begin{equation}
	\bm{S} = \bm{O}^{\mathrm{T}} \mathrm{diag}(s_1, \ldots, s_N) \bm{O} \,,
\end{equation}
where $s_j > 0$, $j = 1, \ldots, N$, and
\begin{equation}
	\bm{O}^{\mathrm{T}} \bm{O} = \bm{O} \bm{O}^{\mathrm{T}} = \uone \,.
\end{equation}
Defining $\bm{\xi} = \bm{O} \bm{x}$, with the inverse $\bm{x} = \bm{O}^{\mathrm{T}} \bm{\xi}$, we have
\begin{equation}
	[ \bm{x} \times \bm{R} \bm{x} ]_j = \sum_{k,l,m} \epsilon_{jkl} x_k R_{lm} x_m = \sum_{k,l,m,n,r} \epsilon_{jkl} O_{nk} \xi_n R_{lm} O_{rm} \xi_r = \sum_{n,r} Q_{jnr} \xi_n \xi_r \,,
\end{equation}
where
\begin{equation}
	Q_{jnr} = \sum_{k,l,m} \epsilon_{jkl} O_{nk} R_{lm} O_{rm} \,.
\end{equation}
Denoting the vector integral in the left-hand side of Eq.~(\ref{eq:app:axial_integral}) by $\bm{\mathcal{I}}$, we write
\begin{equation}
	\mathcal{I}_j = \sum_{n,r} Q_{jnr} \int_{\mathbb{R}^N} d^N \bm{\xi} \, \xi_n \xi_r \exp \left( -\sum_{k} s_k \xi_k^2 \right) = \frac{1}{2} \sqrt{\frac{\pi^N}{\det \bm{S}}} \sum_n \frac{Q_{jnn}}{s_n} \,.
\label{eq:app:intermediata}
\end{equation}
Then,
\begin{equation}
	\sum_n \frac{Q_{jnn}}{s_n} = \sum_{k,l,m} \epsilon_{jkl}  R_{lm} \left( \sum_n O_{nm} \frac{1}{s_n} O_{nk} \right) = \sum_{k,l} \epsilon_{jkl} \big[ \bm{R} \bm{S}^{-1} \big]_{lk} \,.
\end{equation}
Since both $\bm{R}$ and $\bm{S}^{-1}$ are symmetric matrices,
\begin{align}
	\bm{R} \bm{S}^{-1}
	&= \frac{1}{2} \left( \bm{R} \bm{S}^{-1} + \left( \bm{R} \bm{S}^{-1} \right)^{\mathrm{T}} \right) + \frac{1}{2} \left( \bm{R} \bm{S}^{-1} - \left( \bm{R} \bm{S}^{-1} \right)^{\mathrm{T}} \right) \nonumber \\[0.2cm]
	& = \bm{M}_{\mathrm{sym}} + \frac{1}{2} \big[ \bm{R} , \bm{S}^{-1} \big] \,,
\end{align}
where $\bm{M}_{\mathrm{sym}}$ is a symmetric matrix. Consequently, we obtain
\begin{equation}
	\sum_n \frac{Q_{jnn}}{s_n} = \frac{1}{2} \sum_{k,l} \epsilon_{jkl} \big[ \bm{R} , \bm{S}^{-1} \big]_{lk} = -\frac{1}{2} \sum_{k,l} \epsilon_{jkl} \big[ \bm{R} , \bm{S}^{-1} \big]_{kl} = \big[ \axial \big[ \bm{R}, \bm{S}^{-1} \big] \big]_j \,,
\end{equation}
which, combined with Eq.~(\ref{eq:app:intermediata}), yields Eq.~(\ref{eq:app:axial_integral}).

\section{Derivation of Eq.~(\ref{eq:eqs_for_r})}
\label{app:derivatives_of_Phi}

Defining $\bm{\chi}$ in accordance with Eq.~(\ref{eq:chi_def}), and using the fact that $\hat{\bm{\Omega}}$ is symmetric, we have
\begin{align}
	\frac{\partial}{\partial x_j} \bm{\chi}^{\mathrm{T}} \hat{\bm{\Omega}} \bm{\chi}
	&= \frac{\partial}{\partial x_j} \sum_{k,l} \hat{\Omega}_{kl} \chi_k \chi_l \nonumber \\[0.2cm]
	&= \sum_{k,l} \left( \frac{\partial \hat{\Omega}_{kl}}{\partial x_j} \chi_k \chi_l + \hat{\Omega}_{kl} \frac{\partial \chi_k}{\partial x_j} \chi_l + \hat{\Omega}_{kl} \chi_k \frac{\partial \chi_l}{\partial x_j} \right) \nonumber \\[0.2cm]
	&= \bm{\chi}^{\mathrm{T}} \frac{\partial \hat{\bm{\Omega}}}{\partial x_j} \bm{\chi} + 2 \left( \frac{\partial \bm{\chi}}{\partial x_j} \right)^{\mathrm{T}} \hat{\bm{\Omega}} \bm{\chi} \nonumber \\[0.2cm]
	&= \bm{\chi}^{\mathrm{T}} \frac{\partial \hat{\bm{\Omega}}}{\partial x_j} \bm{\chi} + 2 \big[ \hat{\bm{\Omega}} \bm{\chi} \big]_j - 2 \left( \frac{\partial \hat{\bm{q}}}{\partial x_j} \right)^{\mathrm{T}} \hat{\bm{\Omega}} \bm{\chi} \,.
\end{align}
Using Eqs.~(\ref{eq:v-hat}), (\ref{eq:q-hat}), and (\ref{eq:a_and_B}), we write
\begin{equation}
	\frac{\partial \hat{\bm{q}}}{\partial x_j} = -\frac{t^2}{\mu} \frac{\partial \bm{a}_{\bm{x}, \bm{q}}}{\partial x_j} \,,
\label{eq:app:dqdx}
\end{equation}
which leads to
\begin{equation}
	\frac{\partial}{\partial x_j} \bm{\chi}^{\mathrm{T}} \hat{\bm{\Omega}} \bm{\chi} = \bm{\chi}^{\mathrm{T}} \frac{\partial \hat{\bm{\Omega}}}{\partial x_j} \bm{\chi} + 2 \big[ \hat{\bm{\Omega}} \bm{\chi} \big]_j + \frac{2 t^2}{\mu} \left( \frac{\partial \bm{a}_{\bm{x}, \bm{q}}}{\partial x_j} \right)^{\mathrm{T}} \hat{\bm{\Omega}} \bm{\chi} \,.
\label{eq:app:deriv-1}
\end{equation}
Then, in view of Eqs.~(\ref{eq:app:Omega_widetilde}) and (\ref{eq:app:Omega-hat}), evaluation of $\partial \hat{\bm{\Omega}} / \partial x_j$ proceeds as follows:
\begin{align}
	\frac{\partial \hat{\bm{\Omega}}}{\partial x_j}
	&= -\hat{\bm{\Omega}} \, \frac{\partial \hat{\bm{\Omega}}^{-1}}{\partial x_j} \, \hat{\bm{\Omega}} \\[0.2cm]
	&= -\hat{\bm{\Omega}} \, \frac{\partial \widetilde{\bm{\Omega}}^{-1}}{\partial x_j} \, \hat{\bm{\Omega}} \\[0.2cm]
	&= -\frac{t}{\mu} \hat{\bm{\Omega}} \widetilde{\bm{\Omega}}^{-1} \frac{\partial \bm{B}_{\bm{x}, \bm{q}}}{\partial x_j} \widetilde{\bm{\Omega}}^{-1} \hat{\bm{\Omega}} \\[0.2cm]
	&= -\frac{t}{\mu} \hat{\bm{\Omega}} \left( \hat{\bm{\Omega}}^{-1} - \uone t \right) \frac{\partial \bm{B}_{\bm{x}, \bm{q}}}{\partial x_j} \left( \hat{\bm{\Omega}}^{-1} - \uone t \right) \hat{\bm{\Omega}} \\[0.2cm]
	&= -\frac{t}{\mu} \left( \uone - \hat{\bm{\Omega}} t \right) \frac{\partial \bm{B}_{\bm{x}, \bm{q}}}{\partial x_j} \left(  \uone - \hat{\bm{\Omega}} t \right) \,.
\label{eq:app:dOmega_dx}
\end{align}
Substituting Eq.~(\ref{eq:app:dOmega_dx}) into Eq.~(\ref{eq:app:deriv-1}), we obtain
\begin{equation}
	\frac{\partial}{\partial x_j} \bm{\chi}^{\mathrm{T}} \hat{\bm{\Omega}} \bm{\chi} = - \frac{t}{\mu} \bm{\chi}^{\mathrm{T}} \left( \uone - \hat{\bm{\Omega}} t \right) \frac{\partial \bm{B}_{\bm{x}, \bm{q}}}{\partial x_j} \left(  \uone - \hat{\bm{\Omega}} t \right) \bm{\chi} + 2 \big[ \hat{\bm{\Omega}} \bm{\chi} \big]_j + \frac{2 t^2}{\mu} \left( \frac{\partial \bm{a}_{\bm{x}, \bm{q}}}{\partial x_j} \right)^{\mathrm{T}} \hat{\bm{\Omega}} \bm{\chi} \,.
\label{eq:app:deriv-1full}
\end{equation}
Taking the imaginary part, we get
\begin{align}
	\frac{\partial}{\partial x_j} \bm{\chi}^{\mathrm{T}} \hat{\bm{\Omega}}_{\imag} \bm{\chi} = 
	&\frac{t^2}{\mu} \bm{\chi}^{\mathrm{T}} \left[ \hat{\bm{\Omega}}_{\imag} \frac{\partial \bm{B}_{\bm{x}, \bm{q}}}{\partial x_j} \left( \uone - \hat{\bm{\Omega}}_{\real} t \right) + \left( \uone - \hat{\bm{\Omega}}_{\real} t \right) \frac{\partial \bm{B}_{\bm{x}, \bm{q}}}{\partial x_j} \hat{\bm{\Omega}}_{\imag} \right] \bm{\chi} \nonumber \\
	&+ 2 \big[ \hat{\bm{\Omega}}_{\imag} \bm{\chi} \big]_j +  \frac{2 t^2}{\mu} \left( \frac{\partial \bm{a}_{\bm{x}, \bm{q}}}{\partial x_j} \right)^{\mathrm{T}} \hat{\bm{\Omega}}_{\imag} \bm{\chi} \,.
\end{align}
Then, since both $\bm{\Omega}$ and $\bm{B}_{\bm{x}, \bm{q}}$ are symmetric, and since $\bm{\chi}^{\mathrm{T}} \bm{M} \bm{\chi} = \bm{\chi}^{\mathrm{T}} \bm{M}^{\mathrm{T}} \bm{\chi}$ for any matrix $\bm{M}$, the last equality simplifies to
\begin{equation}
	\frac{\partial}{\partial x_j} \bm{\chi}^{\mathrm{T}} \hat{\bm{\Omega}}_{\imag} \bm{\chi} = \frac{2t^2}{\mu} \bm{\chi}^{\mathrm{T}} \left( \uone - \hat{\bm{\Omega}}_{\real} t \right) \frac{\partial \bm{B}_{\bm{x}, \bm{q}}}{\partial x_j} \hat{\bm{\Omega}}_{\imag} \bm{\chi} + 2 \big[ \hat{\bm{\Omega}}_{\imag} \bm{\chi} \big]_j +  \frac{2 t^2}{\mu} \left( \frac{\partial \bm{a}_{\bm{x}, \bm{q}}}{\partial x_j} \right)^{\mathrm{T}} \hat{\bm{\Omega}}_{\imag} \bm{\chi} \,.
\label{app:eq:derivative_chi_Omega_chi}
\end{equation}

We now evaluate the derivative of $\ln \big( \det(\uone + \widetilde{\bm{\Omega}} t) \big)$, with $\widetilde{\bm{\Omega}}$ given by Eq.~(\ref{eq:app:Omega_widetilde}). From the identity
\begin{equation}
\det e^{\bm{M}} = e^{\tr \bm{M}} \,,
\end{equation}
which holds for any complex matrix $\bm{M}$ (see, e.g., Ref.~\cite{Hal03Lie}), it follows that
\begin{equation}
\frac{\partial}{\partial x_j} \ln \left( \det e^{\bm{M}} \right) = \tr \left( \frac{\partial \bm{M}}{\partial x_j} \right) \,.
\end{equation}
Substituting $\bm{M} = \ln (\uone + \widetilde{\bm{\Omega}} t)$, we get
\begin{equation}
\frac{\partial}{\partial x_j} \ln \big( \det (\uone + \widetilde{\bm{\Omega}} t) \big) = \tr \left( (\uone + \widetilde{\bm{\Omega}} t)^{-1} \frac{\partial \widetilde{\bm{\Omega}}}{\partial x_j} t \right) \,,
\end{equation}
which, in view of Eqs.~(\ref{eq:app:Omega_widetilde}) and (\ref{eq:app:tilde-hat-identity}), is equivalent to
\begin{equation}
\frac{\partial}{\partial x_j} \ln \big( \det (\uone + \widetilde{\bm{\Omega}} t) \big) = -\frac{t^2}{\mu} \tr \left[ \left( \uone - \hat{\bm{\Omega}} t \right) \frac{\partial \bm{B}_{\bm{x}, \bm{q}}}{\partial x_j} \right] \,.
\label{eq:app:deriv-4}
\end{equation}
The real part of the last equality reads
\begin{equation}
	\frac{\partial}{\partial x_j} \real \left\{ \ln \left[ \det \left( \uone + \bm{\Omega} t - \frac{1}{\mu} \bm{B}_{\bm{x}, \bm{q}} t^2 \right) \right] \right\} = -\frac{t^2}{\mu} \tr \left[ \left( \uone - \hat{\bm{\Omega}}_{\real} t \right) \frac{\partial \bm{B}_{\bm{x}, \bm{q}}}{\partial x_j} \right] \,.
\label{app:eq:derivative_log_term}
\end{equation}

Finally, substituting Eq.~(\ref{eq:Phi_real}) into Eq.~(\ref{eq:eqs_for_r_def}), and using Eqs.~(\ref{app:eq:derivative_chi_Omega_chi}) and (\ref{app:eq:derivative_log_term}), we arrive at Eq.~(\ref{eq:eqs_for_r}).

\section{Derivation of Eq.~(\ref{eq:Hessian})}
\label{app:derivation_of_Hessian}

Here, we again use the definitions given by Eq.~(\ref{eq:a_and_B}). Partial derivatives of $\bm{\chi}^{\mathrm{T}} \hat{\bm{\Omega}} \bm{\chi}$ and $\ln \left[ \det \left( \uone + \bm{\Omega} t - \frac{1}{\mu} \bm{B}_{\bm{x}, \bm{q}} t^2 \right) \right]$ are given by Eqs.~(\ref{eq:app:deriv-1full}) and (\ref{eq:app:deriv-4}), respectively. We now compute the derivatives of the remaining terms in the right-hand side of Eq.~(\ref{eq:Omega_eff}).

First, we write
\begin{equation*}
	\frac{\partial}{\partial x_j} \hat{\bm{v}}^{\mathrm{T}} \bm{\chi} = \hat{\bm{v}}^{\mathrm{T}} \frac{\partial \bm{\chi}}{\partial x_j} + \left( \frac{\partial \hat{\bm{v}}}{\partial x_j} \right)^{\mathrm{T}} \bm{\chi} = \hat{v}_j - \left( \frac{\partial \hat{\bm{q}}}{\partial x_j} \right)^{\mathrm{T}} \hat{\bm{v}} + \left( \frac{\partial \hat{\bm{v}}}{\partial x_j} \right)^{\mathrm{T}} \bm{\chi} \,.
\end{equation*}
Using Eq.~(\ref{eq:app:dqdx}), along with
\begin{equation}
	\frac{\partial \hat{\bm{v}}}{\partial x_j} = -\frac{t}{\mu} \frac{\partial \bm{a}_{\bm{x}, \bm{q}}}{\partial x_j} \,,
	\label{eq:app:dvdx}
\end{equation}
we get
\begin{equation}
	\frac{\partial}{\partial x_j} \hat{\bm{v}}^{\mathrm{T}} \bm{\chi} = \hat{v}_j + \frac{t^2}{\mu} \left( \frac{\partial \bm{a}_{\bm{x}, \bm{q}}}{\partial x_j} \right)^{\mathrm{T}} \hat{\bm{v}} - \frac{t}{\mu} \left( \frac{\partial \bm{a}_{\bm{x}, \bm{q}}}{\partial x_j} \right)^{\mathrm{T}} \bm{\chi} \,.
	\label{eq:app:deriv-2}
\end{equation}
Second, we have
\begin{equation}
	\frac{\partial}{\partial x_j} \left(  \frac{\mu |\hat{\bm{v}}|^2}{2} - U_{\bm{x}, \bm{q}} \right) = \mu \left( \frac{\partial \hat{\bm{v}}}{\partial x_j} \right)^{\mathrm{T}} \hat{\bm{v}} - \frac{\partial U_{\bm{x}, \bm{q}}}{\partial x_j} \,.
\end{equation}
Using Eq.~(\ref{eq:app:dvdx}), we obtain
\begin{equation}
	\frac{\partial}{\partial x_j} \left(  \frac{\mu |\hat{\bm{v}}|^2}{2} - U_{\bm{x}, \bm{q}} \right) = -t \left( \frac{\partial \bm{a_{\bm{x}, \bm{q}}}}{\partial x_j} \right)^{\mathrm{T}} \hat{\bm{v}} - \frac{\partial U_{\bm{x}, \bm{q}}}{\partial x_j} \,.
	\label{eq:app:deriv-3}
\end{equation}
Then, combining Eqs.~(\ref{eq:app:deriv-1full}), (\ref{eq:app:deriv-4}), (\ref{eq:app:deriv-2}) and (\ref{eq:app:deriv-3}), we find
\begin{align}
	\frac{i \hbar}{2 \mu} \frac{\partial \Phi}{\partial x_j} =
	&\big[ \hat{\bm{\Omega}} \bm{\chi} +  \hat{\bm{v}} \big]_j - \frac{t}{2 \mu} \bm{\chi}^{\mathrm{T}} \left( \uone - \hat{\bm{\Omega}} t \right) \frac{\partial \bm{B}_{\bm{x}, \bm{q}}}{\partial x_j} \left( \uone - \hat{\bm{\Omega}} t \right) \bm{\chi} \nonumber \\
	&-\frac{t}{\mu} \left( \frac{\partial \bm{a_{\bm{x}, \bm{q}}}}{\partial x_j} \right)^{\mathrm{T}} \left( \uone - \hat{\bm{\Omega}} t \right) \bm{\chi} - \frac{t}{\mu} \frac{\partial U_{\bm{x}, \bm{q}}}{\partial x_j} - i \frac{\hbar t^2}{2 \mu^2} \tr \left[ \left( \uone - \hat{\bm{\Omega}} t \right) \frac{\partial \bm{B}_{\bm{x}, \bm{q}}}{\partial x_j} \right] \,.
\label{eq:app:Phi_first_derivative}
\end{align}

We now treat all quantities proportional to the external potential $V$ as being $\epsilon$-small, and differentiate one by one each term in the right-hand side of Eq.~(\ref{eq:app:Phi_first_derivative}) at $\bm{x} = \bm{x}_{\max}$. Using Eqs.~(\ref{eq:chi_def}), (\ref{eq:app:dqdx}) and (\ref{eq:app:dvdx}), we write
\begin{align}
	\frac{\partial}{\partial x_k} \big[ \hat{\bm{\Omega}} \bm{\chi} + \hat{\bm{v}} \big]_j
	&= \hat{\Omega}_{jk} + \left[ \frac{\partial \hat{\bm{\Omega}}}{\partial x_k} \bm{\chi} - \hat{\bm{\Omega}} \frac{\partial \hat{\bm{q}}}{\partial x_k} + \frac{\partial \hat{\bm{v}}}{\partial x_k} \right]_j \nonumber \\[0.2cm]
	&= \hat{\Omega}_{jk} + \left[ \frac{\partial \hat{\bm{\Omega}}}{\partial x_k} \bm{\chi} - \frac{t}{\mu} \left( \uone - \hat{\bm{\Omega}} t \right) \frac{\partial \bm{a}_{\bm{x}, \bm{q}}}{\partial x_k} \right]_j \,.
\label{eq:app:intermediate_derivative_1}
\end{align}
Since
\begin{align}
	\hat{\bm{\Omega}}
	&= \frac{1}{t} \left[ \uone - \left( \uone + \bm{\Omega} t - \frac{1}{\mu} \bm{B}_{\bm{x}, \bm{q}} t^2 \right)^{-1} \right] \nonumber \\[0.2cm]
	&= \frac{1}{t} \left[ \uone - \left( \uone - \frac{t^2}{\mu} (\uone + \bm{\Omega} t)^{-1} \bm{B}_{\bm{x}, \bm{q}} \right)^{-1} (\uone + \bm{\Omega} t)^{-1} \right] \nonumber \\[0.2cm]
	&= \frac{1}{t} \left[ \uone - \left( \uone + \frac{t^2}{\mu} (\uone + \bm{\Omega} t)^{-1} \bm{B}_{\bm{x}, \bm{q}} + \mathcal{O}(\epsilon^2) \right) (\uone + \bm{\Omega} t)^{-1} \right] \nonumber \\[0.2cm]
	&= \frac{1}{t} \left[ \uone - (\uone + \bm{\Omega} t)^{-1} - \frac{t^2}{\mu} (\uone + \bm{\Omega} t)^{-1} \bm{B}_{\bm{x}, \bm{q}} (\uone + \bm{\Omega} t)^{-1} \right] + \mathcal{O}(\epsilon^2) \nonumber \\[0.2cm]
	&= \bm{\Omega}' - \frac{t}{\mu} \invar \bm{B}_{\bm{x}, \bm{q}} \invar + \mathcal{O}(\epsilon^2)
\end{align}
and $\bm{\chi} \big|_{\bm{x} = \bm{x}_{\max}} = \mathcal{O}(\epsilon)$, Eq.~(\ref{eq:app:intermediate_derivative_1}) yields
\begin{equation}
	\left. \frac{\partial}{\partial x_k} \big[ \hat{\bm{\Omega}} \bm{\chi} + \hat{\bm{v}} \big]_j \right|_{\bm{x} = \bm{x}_{\max}} = \left[ \bm{\Omega}' - \frac{t}{\mu} \invar \bm{B}_{\bm{q}', \bm{q}} \invar \right]_{jk} - \frac{t}{\mu} \left[ \invar \frac{\partial \bm{a}_{\bm{q}', \bm{q}}}{\partial q'_k} \right]_j + \mathcal{O}(\epsilon^2) \,.
\label{eq:app:deriv_1_of_5}
\end{equation}
Then,
\begin{equation}
	\left. \frac{\partial}{\partial x_k} \bm{\chi}^{\mathrm{T}} \left( \uone - \hat{\bm{\Omega}} t \right) \frac{\partial \bm{B}_{\bm{x}, \bm{q}}}{\partial x_j} \left( \uone - \hat{\bm{\Omega}} t \right) \bm{\chi} \right|_{\bm{x} = \bm{x}_{\max}} = \mathcal{O}(\epsilon^2) \,,
\label{eq:app:deriv_2_of_5}
\end{equation}
\begin{equation}
	\left. \frac{\partial}{\partial x_k} \left( \frac{\partial \bm{a_{\bm{x}, \bm{q}}}}{\partial x_j} \right)^{\mathrm{T}} \left( \uone - \hat{\bm{\Omega}} t \right) \bm{\chi} \right|_{\bm{x} = \bm{x}_{\max}} = \left[ \left( \frac{\partial \bm{a_{\bm{q}', \bm{q}}}}{\partial q'_j} \right)^{\mathrm{T}} \invar \right]_k + \mathcal{O}(\epsilon^2) = \left[ \invar \frac{\partial \bm{a}_{\bm{q}', \bm{q}}}{\partial q'_j} \right]_k + \mathcal{O}(\epsilon^2) \,,
\label{eq:app:deriv_3_of_5}
\end{equation}
\begin{equation}
	\left. \frac{\partial}{\partial x_k} \frac{\partial U_{\bm{x}, \bm{q}}}{\partial x_j} \right|_{\bm{x} = \bm{x}_{\max}} = \frac{\partial^2 U_{\bm{q}', \bm{q}}}{\partial q'_j \partial q'_k} + \mathcal{O}(\epsilon^2) \,,
\label{eq:app:deriv_4_of_5}
\end{equation}
and
\begin{equation}
	\left. \frac{\partial}{\partial x_k} \tr \left[ \left( \uone - \hat{\bm{\Omega}} t \right) \frac{\partial \bm{B}_{\bm{x}, \bm{q}}}{\partial x_j} \right] \right|_{\bm{x} = \bm{x}_{\max}} = \tr \left[ \invar \frac{\partial^2 \bm{B}_{\bm{q}', \bm{q}}}{\partial q'_j \partial q'_k} \right] + \mathcal{O}(\epsilon^2) = \frac{\partial^2}{\partial q'_j \partial q'_k} \tr \big( \invar \bm{B}_{\bm{q}', \bm{q}} \big) + \mathcal{O}(\epsilon^2) \,.
\label{eq:app:deriv_5_of_5}
\end{equation}
Using Eqs.~(\ref{eq:app:deriv_1_of_5}--\ref{eq:app:deriv_5_of_5}), we obtain
\begin{align}
	\frac{i \hbar}{2 \mu} \frac{\partial^2 \Phi}{\partial x_j \partial x_k} = \Omega'_{jk} - \frac{t}{\mu} \Bigg\{
	&\big[ \invar \bm{B}_{\bm{q}', \bm{q}} \invar \big]_{jk} + \bigg[ \invar \frac{\partial \bm{a}_{\bm{q}', \bm{q}}}{\partial q'_k} \bigg]_j + \bigg[ \invar \frac{\partial \bm{a}_{\bm{q}', \bm{q}}}{\partial q'_j} \bigg]_k \nonumber \\
	&+ \frac{\partial^2 U_{\bm{q}', \bm{q}}}{\partial q'_j \partial q'_k} + i \frac{\hbar t}{2 \mu} \frac{\partial^2}{\partial q'_j \partial q'_k} \tr \big( \invar \bm{B}_{\bm{q}', \bm{q}} \big) \Bigg\} + \mathcal{O}(\epsilon^2) \,.
\label{eq:app:Omega_eff}
\end{align}
In view of the fact that
\begin{equation}
	\bigg[ \invar \frac{\partial \bm{a}_{\bm{q}', \bm{q}}}{\partial q'_k} \bigg]_j = \bigg[ \invar \frac{\partial \bm{a}_{\bm{q}', \bm{q}}}{\partial \bm{q}'} \bigg]_{jk} 
\end{equation}
and
\begin{equation}
	\bigg[ \invar \frac{\partial \bm{a}_{\bm{q}', \bm{q}}}{\partial q'_j} \bigg]_k = \bigg[ \invar \frac{\partial \bm{a}_{\bm{q}', \bm{q}}}{\partial \bm{q}'} \bigg]_{kj} = \left[ \left( \invar \frac{\partial \bm{a}_{\bm{q}', \bm{q}}}{\partial \bm{q}'} \right)^{\mathrm{T}} \right]_{jk} \,,
\end{equation}
Eq.~(\ref{eq:app:Omega_eff}) leads to
\begin{equation}
	\Delta \bm{\Omega} = -\frac{t}{\mu} \bigg[ \invar \bm{B}_{\bm{q}', \bm{q}} \invar + \invar \frac{\partial \bm{a}_{\bm{q}', \bm{q}}}{\partial \bm{q}'} + \left( \invar \frac{\partial \bm{a}_{\bm{q}', \bm{q}}}{\partial \bm{q}'} \right)^{\mathrm{T}} + \frac{\partial^2 U_{\bm{q}', \bm{q}}}{\partial \bm{q}'^2} + i \frac{\hbar t}{2 \mu} \frac{\partial^2}{\partial \bm{q}'^2} \tr \big( \invar \bm{B}_{\bm{q}', \bm{q}} \big) \bigg] \,.
\end{equation}
The last equation is equivalent to Eq.~(\ref{eq:Delta_Omega}).

\twocolumngrid


%

\end{document}